\documentclass[usegraphicx,twocolumn,usenatbib,useAMS]{mn2e} 

\usepackage{mathabx}

\newcommand\Eqn[1]     {Eq.\,(\ref{#1})}

\newcommand\eqn[1]     {Eq.\,(\ref{#1})}
\newcommand\eqns[2]    {Eqs\,(\ref{#1}) and~(\ref{#2})}

\newcommand\fig[1]     {Figure\,{\ref{#1}}}
\newcommand\figs[2]    {Figures\,\ref{#1} and~{\ref{#2}}}
\newcommand\figss[3]     {Figures\,{\ref{#1}}, ~{\ref{#2}} and~{\ref{#3}}}

\newcommand\nn         {\nonumber}

\def\mnras{{Mon.~ Not.~ R.~ Astron.~ Soc.~}}
\def\aj{{Astronomical Journal}}

\def\prd{{Phys.~ Rev.~ D.~}}

\def\apj{{Astrophys.~ J.~}}
\def\apjs{{Astrophys.~ J.~ Suppl.~}}
\def\apjl{{Astrophys.~ J.~ Lett.~}}

\def\mnras{{MNRAS}}

\def\prd{{PRD}}

\def\apj{{ApJ}}
\def\apjs{{ApJS}}
\def\apjl{{ApJL}}
\def\aap{{A\&A}}

\newcommand{\be}{\begin{equation}}
\newcommand{\ee}{\end{equation}}
\newcommand{\ba}{\begin{eqnarray}}
\newcommand{\ea}{\end{eqnarray}}

\newcommand{\mx}{\mbox}
\newcommand{\bm}{\boldmath }
\newcommand{\euclid}{\it{Euclid}}
\newcommand{\lsst}{\textsc{lsst}}

\newcommand{\planck}{\textsc{planck}}
\newcommand{\arcmint}{\mathrm{arcmin}}
\newcommand{\zmed}{z_{\mathrm{med}}}
\newcommand{\thetapix}{\theta_{\mathrm{pix}}}
\newcommand{\arcsect}{\mathrm{arcsec}}
\newcommand{\C}{\mbox{\boldmath $C$}}

\newcommand{\p}{\mbox{\boldmath $p$}}
\newcommand{\m}{\mbox{\boldmath $m$}}
\newcommand{\map}{M_{\rm ap}}

\newcommand{\corra}{\omega_{\rm pp}^{\rm a}}
\newcommand{\corrc}{\omega_{\rm pp}^{\rm c}}

\def\SN{{\mathcal S}/{\mathcal N}}
\def\Fish{{\mathcal F}}

\def\btheta{{\mx {\bm $\theta$}}}

\def\km{{\kappa_{\rm m}}}
\def\kmsq{{\kappa^2_{\rm m}}}
\def\bkm{{\bar\kappa_{\rm m}}}

\def\bkmsq{{\bar\kappa^2_{\rm m}}}

\def\Mm{{M_{\rm m}}}

\def\map{{M_{\rm ap}}}

\def\deg2{\rm deg^2}
\def\arcmin2{\rm arcmin^2}

\def\nn{{\nonumber}}
\def\om{{\Omega_{\rm m}}}
\def\s8{{\sigma_8}}
\def\ns{{n_{\rm s}}}

\def\Hub{{\rm km}s^{-1}{\rm Mpc}^{-1}}

\def\kpc{\, h^{-1}{\rm kpc}}
\def\Mpc{\, h^{-1}{\rm Mpc}}

\def\nbar{\bar{n}}

\title
{The cosmological information of shear peaks: beyond the abundance }

\author[Marian, Smith, Hilbert \& Schneider] { Laura
  Marian,$^{1,2}$\thanks{lmarian@MPA-Garching.MPG.DE}
  Robert~E.~Smith,$^{1}$ Stefan Hilbert$^{1,3}$ and Peter
  Schneider$^{2}$ \\ $^2$ Argelander-Institute for Astronomy, Auf dem
  H\"ugel 71, D-53121 Bonn, Germany\\ $^1$ Max-Planck Institute for
  Astrophysics, Karl-Schwarzschild-Str. 1, Garching, D-85748\\ $^3$ Kavli
Institute of Particle Astrophysics and Cosmology (KIPAC), Stanford
University, 452 Lomita Mall, Stanford, CA 94305}

\begin{document}

\voffset -1.5cm

\maketitle


\begin{abstract}
We study the cosmological information of weak lensing (WL) peaks,
focusing on two other statistics besides their abundance: the stacked
tangential-shear profiles and the peak-peak correlation function. We
use a large ensemble of simulated WL maps with survey specifications
relevant to future missions like $\euclid$ and $\lsst$, to measure and
examine the three peak probes. We find that the auto-correlation
function of peaks with high signal-to-noise ($\SN$) ratio measured
from fields of size 144 $\deg2$ has a maximum of $\sim 0.3$ at an
angular scale $\vartheta\sim 10$ arcmin. For peaks with smaller $\SN$,
the amplitude of the correlation function decreases, and its maximum
occurs on smaller angular scales. The stacked tangential-shear
profiles of the peaks also increase with their $\SN$. We compare the
peak observables measured with and without shape noise and find that
for $\SN\sim3$ only $\sim5\%$ of the peaks are due to large-scale
structures, the rest being generated by shape noise. The correlation
function of these small peaks is therefore very weak compared to that
of small peaks measured from noise-free maps, and also their mean
tangential-shear profile is a factor of a few smaller than the
noise-free one. The covariance matrix of the probes is examined: the
correlation function is only weakly covariant on scales $\vartheta<30$
arcmin, and slightly more on larger scales; the shear profiles are
very correlated for $\vartheta>2$ arcmin, with a correlation
coefficient as high as 0.7. The cross-covariance of the three probes
is relatively weak: the peak abundance and profiles have the largest
correlation coefficient $\sim 0.3$. Using the Fisher-matrix formalism,
we compute the cosmological constraints for $\{\om,\,\s8,\,w,\,\ns\}$
considering each probe separately, as well as in combination. We find
that the peak-peak correlation and shear profiles yield marginalized
errors which are larger by a factor of $2-4$ for $\{\om,\,\s8\}$ than
the errors yielded by the peak abundance alone, while the errors for
$\{w,\,\ns\}$ are similar. By combining the three probes, the
marginalized constraints are tightened by a factor of $\sim 2$
compared to the peak abundance alone, the least contributor to the
error reduction being the correlation function. This work therefore
recommends that future WL surveys use shear peaks beyond their
abundance in order to constrain the cosmological model.
\end{abstract}

\section{Introduction}
\label{S1}
With the approach of large weak gravitational lensing (WL) missions,
such as $\euclid$ and $\lsst$ \citep{lsst2009, Euclid2011}, it is
imperative to optimize the extraction of cosmological information from
WL probes. Until now, the WL community has been using mainly the
2-point function of the shear field in order to obtain cosmological
constraints \citep{Jarvisetal2003, Hoekstraetal2006,
  Sembolonietal2006, Hetterscheidtetal2007, Kilbingeretal2012}, and
also to plan and forecast coming missions \citep{Euclid2011}. Whilst a
lot of effort is concentrated on scrutinizing the systematics of WL
surveys -- a most essential endeavour for the success of the
next-generation missions, and indeed for the future of WL as a
cosmology probe -- the task of selecting the most efficient statistics
for parameter estimation is also a crucial aspect that should not be
neglected.

One of the possible WL observables involves the `peaks' of the shear
field, i.e. regions of high signal-to-noise ($\SN$) of the field,
produced by overdense regions of the density field projected along the
line of sight. Shear peaks are the imprint of clusters in WL maps, and
can be used to detect and measure cluster masses
\citep{Hamanaetal2004, Wangetal2004, Maturietal2005,
  HennawiSpergel2005, Dahle2006, MarianBernstein2006,
  Schirmeretal2007, Maturietal2007, Abateetal2009}. The peak abundance
scales with cosmological parameters in the same way as the halo mass
function \citep{Reblinskyetal1999, Marianetal2009, Marianetal2010},
and therefore can be used to constrain the cosmological model
\citep{DietrichHartlap2010, Kratochviletal2010, Marianetal2012,
  Bardetal2013}. The shear-peak abundance can also constrain
primordial non-Gaussianity \citep{Marianetal2011, Maturietal2011,
  Hilbertetal2012}, and so far seems to be possibly the most effective
WL observable suitable for that purpose -- although more work must be
done on this subject.

Whilst the abundance of peaks has been investigated by a number of
studies, higher-order statistics of the shear peaks have until now
been overlooked. Yet, since the abundance of peaks can be used as a
cosmological tool, one would expect their clustering to also be
valuable. In analogy with 3D observables, just as both the halo
correlation function and the halo mass function are very sensitive to
cosmology, the same could be true about the shear peaks.

The main goal of this study is to investigate the correlation function
of WL peaks, and in particular to quantify the improvements that it
yields on cosmological constraints relative to the peak abundance. As
a secondary line of inquiry, we shall also pursue the cosmological
information contained in the tangential-shear profiles of the peaks.
While these statistics -- correlations and stacked profiles of
overdensities -- have been already studied in various theoretical
works, as well as by using real and simulated data \citep{
  Sheldonetal2004, Sheldonetal2009, Mandelbaumetal2010,
  OguriTakada2011, Mandelbaumetal2012, Cacciatoetal2012}, here we
explore the characteristics of shear-selected overdensities for which
the only known information is the redshift distribution of the source
galaxies. This might seem a somewhat unlikely and conservative
scenario for future surveys, but here we just examine the intrinsic
information contained in the peaks, even if our results will be
sub-optimal on this account. We defer to a near-future study a more
realistic scenario, where prior information on the peaks is available,
and where we can account for various survey uncertainties, such as
intrinsic alignments, masked regions, calibration errors of the source
galaxies, etc.

We shall present measurements of the peak function, peak-peak
correlation function, and peak profiles from an ensemble of WL maps
generated with ray-tracing through $N$-body simulations with varying
cosmologies, as described in \S\ref{S2}. We shall use the
Fisher-matrix formalism to establish the cosmological sensitivity of
the above-mentioned probes. We underscore that our results are drawn
from collaborative efforts, which involved generating the $N$-body
simulations, the ray-tracing maps, developing the hierarchical
algorithm for WL peaks; the present study is part of a larger research
program, which has also yielded the results published in
\cite{Marianetal2011, Marianetal2012} and
\cite{Hilbertetal2012}. Therefore, we shall not give a full
description of all our tools, but simply refer to these publications.

The paper is partitioned as follows: in \S\ref{S2}, we provide a
description of the ray-tracing maps used in this study; in \S\ref{S3}
we illustrate our measurements of the three peak probes; in \S\ref{S4}
we present the cosmological constraints extracted from the three
probes; in \S\ref{S5} we summarize and conclude.
\section{Simulated WL maps}
\label{S2}
We generated WL maps from ray-tracing through $N$-body simulations. We
used 8 simulations which are part of a larger suite performed on the
zBOX-2 and \mbox{zBOX-3} supercomputers at the University of
Z\"{u}rich. For all realizations 11 snapshots were output between
redshifts $z=[0,2]$; further snapshots are at redshifts
$z=\{3,4,5\}$. We shall refer to these simulations as the {\tt
  zHORIZON} simulations, and they were described in detail in
\cite{Smith2009, Smithetal2012}.  Each of the {\tt zHORIZON}
simulations was performed using the publicly available {\tt Gadget-2}
code \citep{Springel2005}, and followed the nonlinear evolution under
gravity of $N=750^3$ equal-mass particles in a comoving cube of length
$L_{\rm sim}=1500\Mpc$; the softening length was $l_{\rm soft}=60\,
\kpc$. The cosmological model was similar to that determined by the
WMAP experiment \citep{Komatsuetal2009short}. We refer to this
cosmology as the fiducial model. The transfer function for the
simulations was generated using the publicly available {\tt cmbfast}
code \citep{SeljakZaldarriaga1996}, with high sampling of the spatial
frequencies on large scales. Initial conditions were set at redshift
$z=50$ using the serial version of the publicly available {\tt 2LPT}
code
\citep{Scoccimarro1998,Crocceetal2006}. Table~\ref{tab:zHORIZONcospar}
summarizes the cosmological parameters that we simulated. We also use
another series of simulations, identical in every way to the fiducial
model, except that we have varied one of the cosmological parameters
by a small amount. For each new set we have generated 4 simulations,
matching the random realization of the initial Gaussian field with the
corresponding one from the fiducial model. The four parameter
variations were: $\{n\rightarrow (0.95, 1.05 ),\,\s8\rightarrow (0.7,
0.9),\,\om\rightarrow (0.2, 0.3),\, w\rightarrow (-1.2, -0.8) \}$, and
we refer to each of the sets as {\tt zHORIZON-V1a,b},\dots,{\tt
  zHORIZON-V4a,b}, respectively. 

For the WL simulations, we considered a survey similar to $\euclid$
\citep{Euclid2011} and to \lsst{} \citep{lsst2009}, with: an rms
$\sigma_{\gamma}=0.3$ for the intrinsic image ellipticity, a source
number density $\nbar =40\,\arcmint^{-2}$, and a redshift distribution
of source galaxies given by
$
{\mathcal P}(z)={\cal N}(z_0, \beta)\,z^2\exp[-(z/z_0)^{\beta}], 
$
%
\begin{table*}
\caption{{\tt zHORIZON} cosmological parameters. Columns
are: density parameters for matter, dark energy and baryons; the
equation of state parameter for the dark energy;
normalization and primordial spectral index of the power spectrum;
dimensionless Hubble parameter.
\label{tab:zHORIZONcospar}}
\vspace{0.2cm}
\centering{
\begin{tabular}{c|ccccccc}
\hline 
Cosmological parameters & $\om$ & $\Omega_{\rm DE}$ & $\Omega_b$ & $w$  &  
$\s8$  & $n$ &  $H_{0} [\Hub]$ \\
\hline
{\tt zHORIZON-I      }  & 0.25\ &  0.75 & 0.04 &  -1  &  0.8  & 1.0 & 70.0\\
{\tt zHORIZON-V1a/V1b}  & 0.25\ &  0.75 & 0.04 &  -1  &  0.8  & 0.95/1.05 & 70.0\\
{\tt zHORIZON-V2a/V2b}  & 0.25\ &  0.75 & 0.04 &  -1  &  0.7/0.9  & 1.0 & 70.0\\
{\tt zHORIZON-V3a/V3b}  & 0.2/0.3\ &  0.8/0.7 & 0.04 &  -1  &  0.8  & 1.0 & 70.0\\
{\tt zHORIZON-V4a/V4b}  & 0.25\ &  0.75 & 0.04 &  -1.2/-0.8  &  0.8  & 1.0 & 70.0\\
\end{tabular}}
\end{table*}
%
where the normalization constant ${\cal N}$ ensures that the integral
of the source distribution over the source redshift is unity. If this
interval extended to infinity, then the normalization could be written
analytically as: ${\cal N}=3/(z_0^3\,\Gamma[(3+\beta)/\beta])$. There
is a small difference between this value and what we actually used,
due to the fact that we considered a source interval of $[0, 3]$. We
took $\beta=1.5$, and required that the median redshift of this
distribution be $\zmed=0.9$, which fixed $z_0\approx 0.64$, and gave a
mean of $z_{\rm mean}=0.95$.
\par From each $N$-body simulation we generated 16 independent fields of
view. Each field had an area of $12 \times 12\,\deg2$ and was tiled by
$4096^2$ pixels, yielding an angular resolution $\thetapix
=10\,\arcsect$. For each variational model, the total area was of
$\approx 9000\, \deg2$, while for the fiducial model it was of
$\approx 18000\, \deg2$. The effective convergence $\kappa$ in each
pixel was calculated by tracing a light ray back through the
simulation with a multiple-lens-plane ray-tracing algorithm
\citep{Hilbertetal2007, Hilbertetal2009}. Gaussian shape noise with
variance $\sigma_{\gamma}^2 /(\nbar\, \thetapix^{2})$ was then added
to each pixel, creating a realistic noise level and correlation in the
filtered convergence field \citep{HilbertMetcalfWhite2007}. We keep
the shape noise configuration fixed for each field in different
cosmologies, in order to minimize its impact on the comparisons of the
peak abundances measured for each cosmology.

\section{WL peaks as a cosmological probe}
\label{S3}
As stated in  \S\ref{S1}, we shall consider the cosmological
  information contained in three shear peak probes: the abundance, the
  profiles, and the correlation function. Until now, only the
  abundance of peaks has been examined as a probe for cosmology; in
  this study we shall investigate the cosmological information
  contained in the clustering of peaks and their profiles, as well as
  various combinations of the three probes.

The peaks are detected with an aperture-mass filter, i.e. a
compensated filter \cite[][]{Schneider1996}, as points of maxima in the
smoothed convergence field. They are assigned a unique $\SN$ and mass
using the hierarchical algorithm. This method, as well as the
NFW-shaped filter function, were described in detail in our previous
work \cite[][]{Marianetal2012}. The hierarchical algorithm uses a
hierarchy of filters of different size, from the largest down to the
smallest, to determine the size of the NFW filter that best matches
each peak. For every filter used, we write the aperture mass and the
$\SN$ at a given point $\btheta_0$ as
\be \map(\btheta_0) =
\Mm\frac{\int d^2\theta\,[\km(\theta)- \bkm(\theta_{\rm
      A})]\kappa(\btheta_0-\btheta)}{\int d^2\theta \,
  \kmsq(\theta)-\pi\theta^2_{A}\bkmsq(\theta_{A})},
\label{eq:smoothed_map2}
\ee
\be
\SN(\btheta_0)=\sqrt{\frac{\nbar}{\sigma_{\gamma}^2}}\, \frac{\int d^2\theta\,
  [\km(\theta)-\bkm(\theta_{\rm A})]\kappa(\btheta_0-\btheta)} 
     {\sqrt{\int d^2\theta\,[\km(\theta)-\bkm(\theta_{\rm A})]^2}},
\label{eq:SNc}
\ee
where $\Mm$ is the mass of the NFW halo used as a model for the
filter, and $\theta_{\rm A}$ is the aperture radius of the filter,
which in our case was chosen to be the angular size of its NFW
radius. $\km$ is the convergence profile of the model, $\bkm$ is the
mean convergence inside a certain radius, and $\kappa$ is the measured
convergence field. If $\btheta_0$ is the location of a peak generated
by an NFW halo matching the filter, then the above equations become
\be \map(\btheta_0) = \Mm,
\label{eq:Mmax}
\ee
\be \SN(\btheta_0)=\sqrt{\frac{\nbar}{\sigma_{\gamma}^2}\,\int
  d^2\theta\,[\km(\theta)-\bkm(\theta_{\rm A})]^2}.
\label{eq:SNmax}
\ee
The hierarchical method enhances the detection of peaks, their
abundance as a function of mass or $\SN$ is independent of any
particular choice of filter size, and the cosmological constraints
arising from the peak counts are tighter compared to the case when a
single-sized filter is used. It is also instructive to present
measurements of the peak function, the profiles and the peak-peak
correlation function from noise-free maps. In that case, the peaks are
selected based on their aperture-mass values, and they are assigned a
hierarchical mass through \eqn{eq:Mmax}, as described in
\cite{Marianetal2012}. For the purpose of comparison with the results
from the noisy maps, we can assign the noise-free peaks a fictitious
$\SN$ value, using \eqns{eq:Mmax}{eq:SNmax} with the same level of
shape noise as in the noisy maps.

\subsection { The peak-peak correlation function }
\begin{figure*}
\centering {
\includegraphics[scale=0.44]{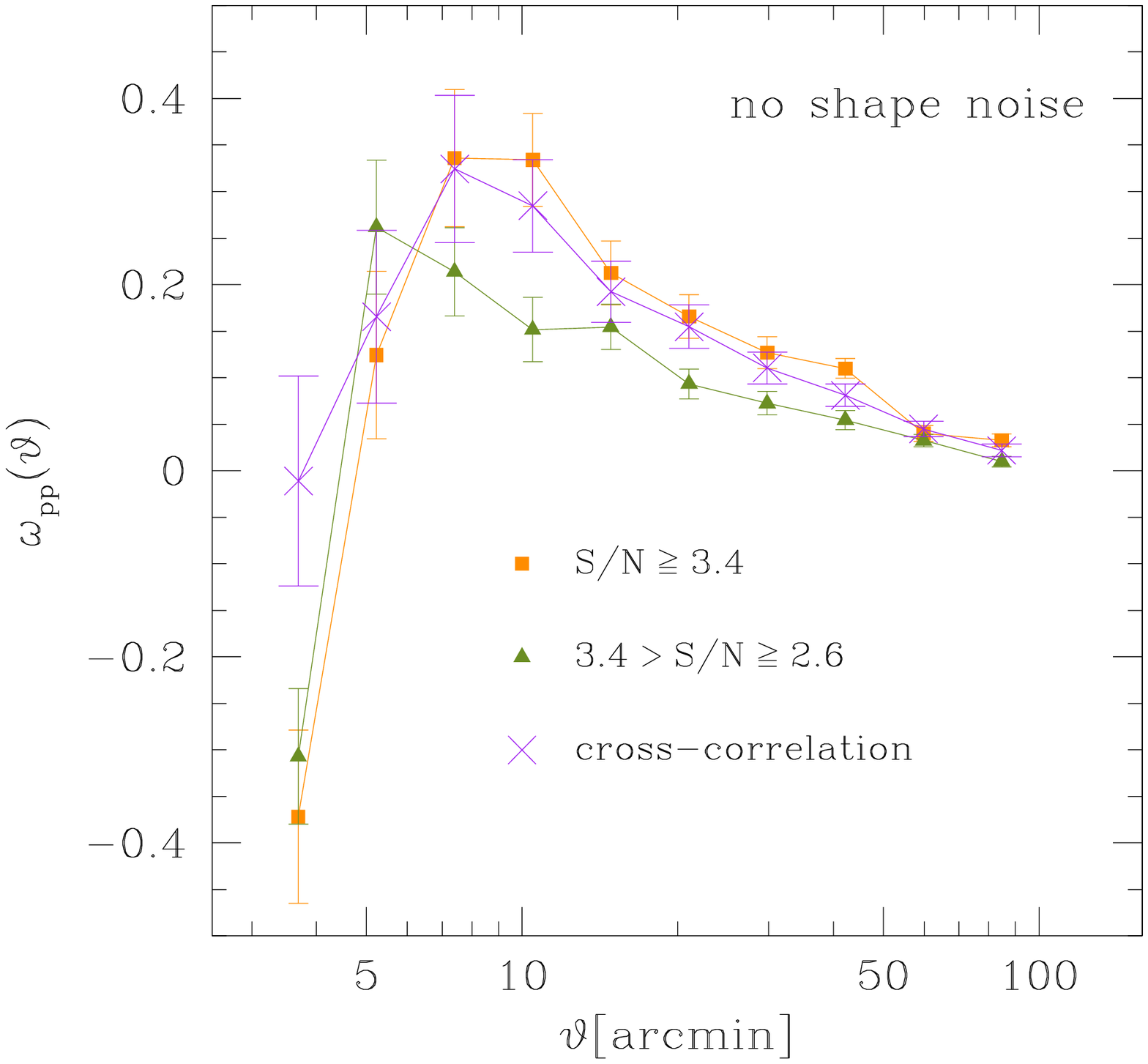}
\includegraphics[scale=0.44]{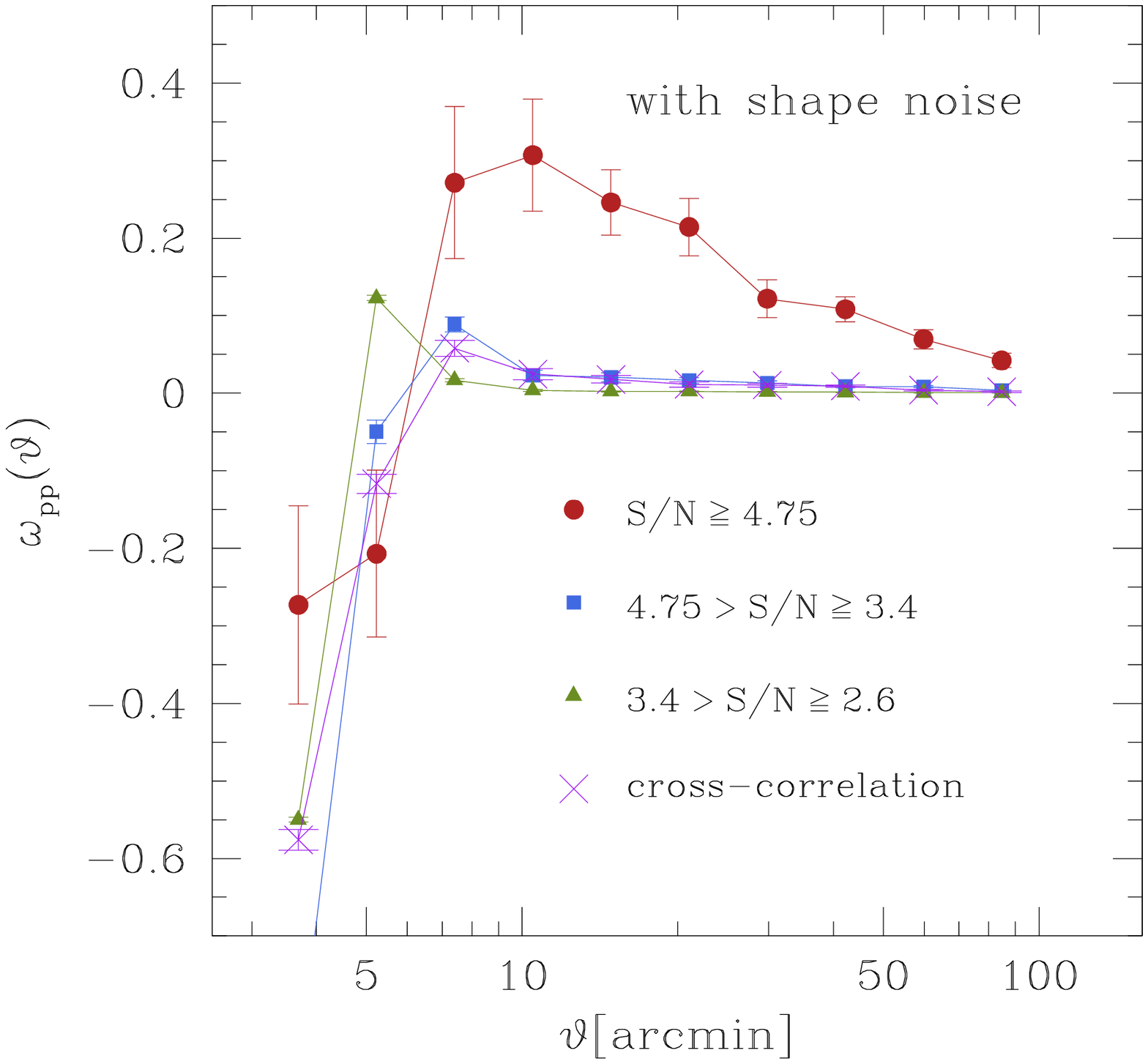}}
\caption{The correlation function of the peaks detected in the maps
  corresponding to the fiducial cosmological model. The results show
  the average of 128 fields of $12\times12\,\deg2$ and the error bars
  represent errors on the mean. {\it Left panel:} The orange
  squares/green triangles denote measurements in the absence of shape
  noise, for peaks with $\SN\geq3.4$, and $3.4>\SN\geq2.6$
  respectively. {\it Right panel:} Measurements from the noisy maps:
  the solid red circles, blue squares, green triangles represent the
  auto-correlation of peaks with $\SN\geq4.75$, $4.75>\SN\geq3.4$,
  $3.4>\SN\geq2.6$ respectively. In both panels, the purple crosses
  depict the cross-correlation of the peak populations with
  $3.4>\SN\geq2.6$ and $\SN\geq4.75$, where we have used
  \eqns{eq:Mmax}{eq:SNmax} to assign an $\SN$ value to the peaks.}
\label{fig:corrfunc}
\end{figure*}
\begin{figure*}
\centering {
\includegraphics[scale=0.44]{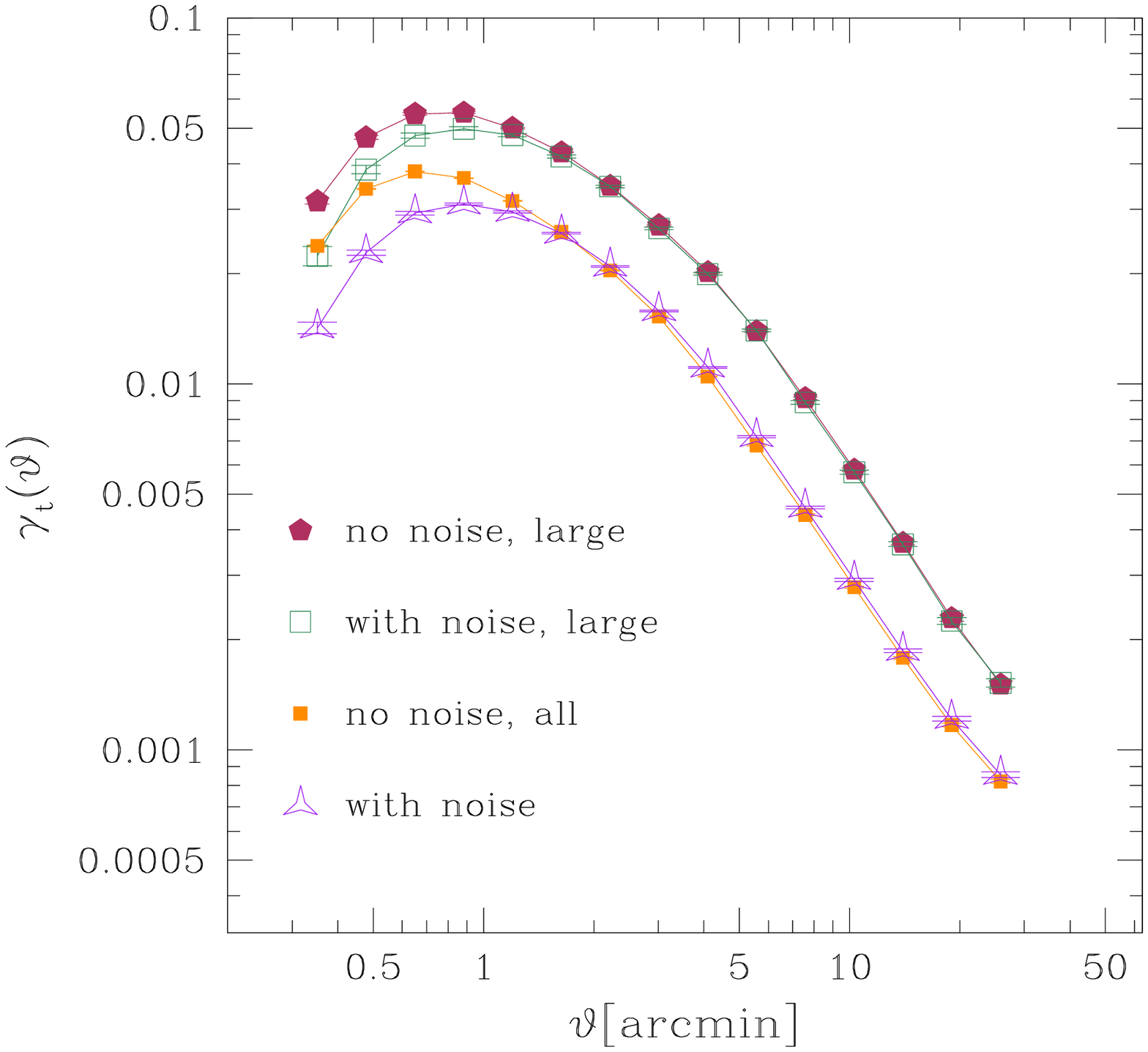}
\includegraphics[scale=0.44]{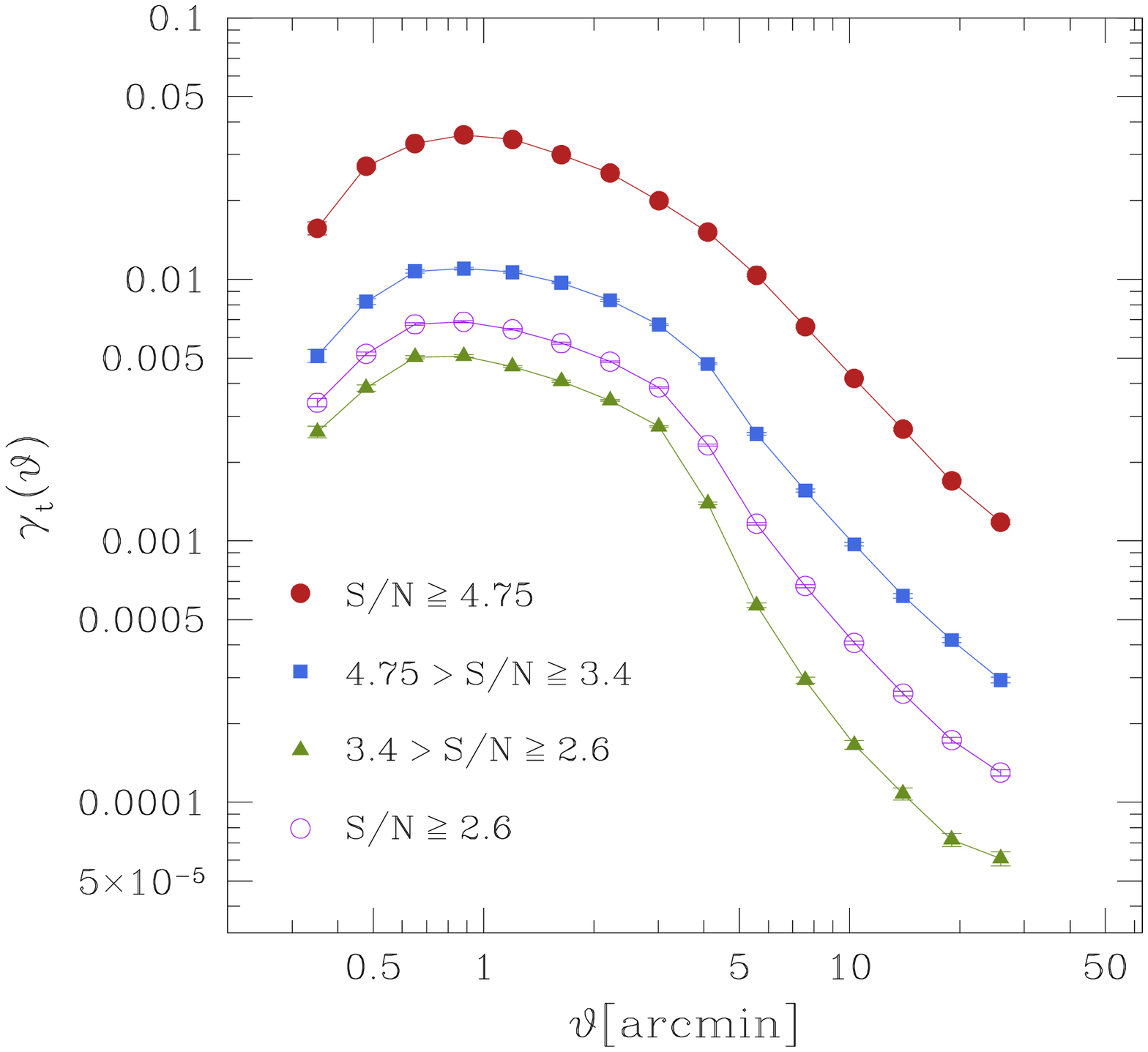}}
\caption{Tangential-shear profiles for the peaks detected in the maps
  corresponding to the fiducial cosmological model. {\it Left panel:}
  The effect of shape noise on the profiles. The red solid pentagons
  represent peaks detected in the noise-free maps, with
  $\SN\geq4.75$. The green empty squares depict the profiles of these
  very same peaks, detected in the noisy maps. Similarly, the orange
  solid squares show the profiles of all the peaks that we detected in
  the noise-free maps, while the purple stars show the profiles of the
  same peaks when detected and measured in the noisy maps. {\it Right
    panel:} The profiles measured from the noisy maps: the solid red
  circles, blue squares, and green triangles represent peaks with
  $\SN$ indicated in the legend. The purple circles depict all the
  measured peaks binned together. }
\label{fig:profiles}
\end{figure*}
We have measured the correlation function of the peaks using the
minimum-variance and unbiased estimator introduced by
\cite{LandySzalay1993}. The estimator is:
\be
\hat\omega_{\rm pp}=(DD-2DR+RR)/RR,
\ee
where $DD$, $DR$, and $RR$ represent the peak-peak, peak-random peak,
and random peak-random peak pair counts, respectively. The correlation
function is measured in 10 angular bins, logarithmically spaced
between $\vartheta\in\left[4, 90\right]$ arcmin. The lower bound
was chosen to be the smallest possible without discreteness effects
contaminating the measurements -- the latter appear when the size of
the lower bins is only a small multiple of the size of the pixels of
the map. The random catalogue that we created in order to build the
correlation function estimator contained a million random peaks placed
in a field of $12\times12\,\deg2$, i.e. the same size as the simulated
maps.

In \fig{fig:corrfunc} we present measurements of the correlation
function of peaks detected from the maps corresponding to the fiducial
cosmology. All symbols denote the average of 128 realizations, the
errors being on the mean. The estimator of the average correlation
function is defined by \eqn{eq:estmean}; a similar estimator will also
be used to obtain the average tangential-shear profile of the peaks
and the average peak function presented later in this section.

The left panel of \fig{fig:corrfunc} shows the noise-free results,
while in the right one shape noise is included. The solid orange
squares/green triangles in the left panel depict the auto-correlation
$\corra$ of the peaks with $3.4\geq\SN\geq2.6$ and $\SN\geq3.4$,
respectively. The solid red circles, blue squares, and green triangles
in the right panel correspond to peaks with $\SN\geq 4.75$,
$4.75\geq\SN\geq3.4$, and $3.4\geq\SN\geq2.6$ respectively. The total
number of peaks in these bins is $\sim 10,000,\,72,000,\,324,000$
respectively. In both panels, the violet crosses depict the
cross-correlation $\corrc$ of the bins $3.4\geq\SN\geq2.6$ and
$\SN\geq 4.75$.

The correlation functions are strongly negative for the smallest bins
due to an exclusion effect arising from our peak selection: we discard
those peaks whose centres are separated from the centre of a larger
peak by a distance smaller than approximately one virial radius. The
reason for this choice is to avoid classifying substructures as
independent peaks.

\begin{figure}
\centering {
\includegraphics[scale=0.44]{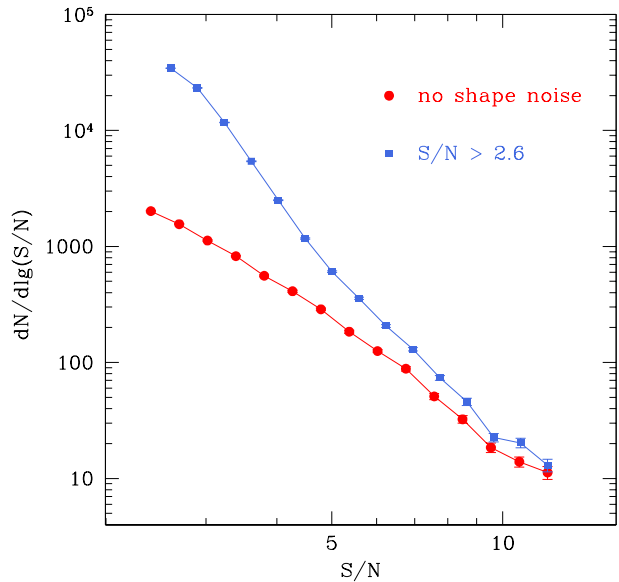}}
\caption{The peak function measured from the maps corresponding to the
  fiducial cosmological model. The results show the average of 128
  fields of 144 $\deg2$. The red points denote measurements in the
  absence of shape noise; the blue squares represent measurements from
  noisy maps, with an $\SN$-threshold of 2.6. }
\label{fig:peakfunc}
\end{figure}

In general, the peaks with higher/lower $\SN$ have higher/lower
correlations, similar to the clustering properties displayed by 3D
halos. For the noiseless measurements, the maximum of the
auto-correlation of peaks with $\SN\geq3.4$ is reached at $\sim
8\,\arcmint$ and is about 0.35, while for the peaks with
$3.4\geq\SN\geq2.6$ the maximum is about 0.25 and it is shifted
towards smaller scales $\sim 5\,\arcmint$. The auto-correlation
functions decrease monotonically to $0$ with increasing angular
scales. The same pattern is followed by the correlation functions of
the peaks measured from the noisy maps: the higher the $\SN$ bin, the
larger the amplitude of the function and the angular position of its
maximum. For the largest peaks, a maximum of $\sim 0.3$ is reached at
$\sim 10$ arcmin. Comparing the measurements from noise-free and noisy
maps allows us to understand how shape noise impacts the correlation
function of the peaks. For example, the green triangles in the left
and right panels of \fig{fig:corrfunc} are dissimilar, although peaks
from the same $\SN$ bin are used for the measurements. The noise-free
correlation function is significantly higher, dropping to 0 only at
$\sim100\,\arcmint$, unlike its noisy counterpart, which vanishes at
$\sim8\,\arcmint$. The vanishing owes to the fact that most peaks in
this $\SN$ bin are generated by shape noise -- see also
\fig{fig:peakfunc} -- and behave essentially like random points. The
genuine signal seen in the left panel of the figure is `drowned' by
the large number of shape-noise pair counts. The same is true for the
cross-correlation signal depicted by the purple crosses in both
panels. The noise-free signal is quite high, similar to the
auto-correlation of peaks with $\SN\geq3.4$, and the error bars are
significantly smaller than for the auto-correlation of the peaks with
$\SN\geq 4.75$ -- high-$\SN$ peaks are very scarce in the noise-free
maps, hence the errors on their correlation function are quite large,
which is the reason why we do not even show it in the
figure. Regarding the noisy maps, whilst the cross-correlation
function has a smaller signal-to-noise than the auto-correlation of
the largest peaks, it can be used to constrain cosmology, as shown in
\S\ref{S4}.
\subsection{The tangential-shear profiles of the peaks}
We measure the tangential-shear profiles $\gamma_{T}$ around the
centres of the peaks, as found through our filtering method. We use 15
angular bins, logarithmically spanning the interval $\vartheta\in[0.3,
  26]\,\arcmint$. In \fig{fig:profiles} we present the average of the
stacked profiles, measured from 128 fields corresponding to the
fiducial cosmology.

The left panel shows the impact of shape noise on the peak
profiles. The solid red pentagons and orange squares depict the
stacked profiles of the largest peaks ($\SN\geq4.75$), and all the
peaks identified in the noise-free maps. The number of peaks
contributing to the two curves are $\sim 4,500$ and $\sim 39,000$
respectively. The noise-free peaks are generated by lensing of
large-scale structures (LSS); therefore, they are also present in the
noisy maps, albeit with slightly different centre coordinates and
amplitude. We match the coordinates of the noise-free peaks to the
coordinates of the peaks in the noisy maps. This is done on a
field-by-field basis, allowing for a displacement of 5 pixels around
the centre of the noise-free peaks. The green empty squares and purple
stars indicate the stacked profiles of the same peaks depicted by the
solid red pentagons and orange squares, when measured from the noisy
maps. The shape noise renders the measured profiles shallower on small
scales, $\vartheta\leq\,2\,\arcmint$, due to the shift in the centre
coordinates. On scales $\vartheta > \,2\,\arcmint$ there is no
significant difference between the noise-free and noisy profiles.
The right panel shows measurements from the noisy maps. The $\SN$-bins
are similar to those used for the correlation function: $3.4>\SN\geq
2.6$ -- green solid triangles; $4.75>\SN\geq 3.4$ -- blue solid
squares; and $\SN\geq 4.75$ -- red solid circles. For illustration
purposes, we also include the stacked profiles of all peaks binned
together, represented by the purple empty circles. The more massive
the peaks, the higher their profiles: there is about an order of
magnitude difference between the average tangential shear
corresponding to the highest and lowest $\SN$-bins. Similar to the
correlation function, the stacked profiles of the noisy peaks are
significantly lower than the noiseless ones. For example, compare the
average profile of all peaks from the two panels, i.e. the orange
solid squares on the left with the empty purple circles on the right:
there is a factor of $\sim\,5$ difference on small scales, and less on
larger scales. This is a consequence of several factors: i) most of
the low-$\SN$ peaks are caused by shape noise and they represent
random points in the shear maps, thereby lowering the profile average
of the real peaks; ii) some small genuine LSS peaks are boosted by
shape noise above the detection threshold, and so they are included in
the right panel, but not in the left one; iii) as shown in the left
panel, shape noise does tend to lower the average profiles due to the
shifting of the peak centre.
\subsection{The peak function}
\fig{fig:peakfunc} illustrates the peak function $\Phi$, i.e. the
number of peaks per unit $\SN$, in a field with area 144 $\deg2$. We
considered 15 logarithmic $\SN$ bins, in the interval $[2.6, 14]$. The
red solid circles denote the noiseless peak function measured from 128
fields corresponding to the fiducial cosmology, while the blue solid
squares are the same measurement in the presence of shape noise. The
two peak functions are very similar at the high-$\SN$ end, and differ
by a factor of at least 20 at the low-$\SN$ end, where peaks generated
by shape-noise completely dominate the genuine LSS peaks. Modelling
the effect of shape noise on the peak function is the subject of a
work in progress, so we shall not dwell on it any longer here. We
merely mention that the dominance of shape-noise peaks at the
low-$\SN$ end of the peak function is in agreement with the behaviour
of the peak-peak correlation function, and the tangential-shear
profiles displayed in \figs{fig:corrfunc}{fig:profiles}.
\section{Fisher matrix calculations}
\label{S4}
In this section we employ the Fisher-matrix formalism to derive
cosmological constraints from the three peak probes presented in the
preceding section. We shall make use of the cosmology dependence of
our simulated maps, and measure the peak statistics corresponding to
the models varying cosmological parameters around the fiducial
values. These measurements can then be used to obtain Fisher
predictions.

Throughout this work we shall assume the likelihood function of the
peak probes to be a multivariate Gaussian. Consider a vector of
measurements of the peak-peak correlation function, peak abundance,
and peak profiles $\m=\{\omega_{1},\dots, \omega_{n_{\rm B}^\omega},
\Phi_1 \dots, \Phi_{n_{\rm B}^\Phi}, \gamma_{T}^{1}, \dots,
\gamma_{T}^{\rm n_{\rm B}^{\gamma_T}}\}$, where $n_{\rm
  B}^{\omega},\,n_{\rm B}^{\Phi},\,n_{\rm B}^{\gamma_T}$ are the
number of bins for the respective measurements. If we similarly define
a vector $\bar{\m}$ of the mean of the measurements, we write the
likelihood as
\ba {\cal L}(\m|\,\bar{\m}(\p), \C(\p))=\frac{1}{(2\pi)^{n_{\rm
      B}/2}|\C|^{1/2}}
\hspace{2.6cm} \nn \\ \times
\exp\left[-\frac{1}{2}(\m-\bar{\m})^{t}\C^{-1} (\m-\bar{\m})\right],
\ea
where the variable $\p$ indicates the dependence on the cosmological
model of the mean and covariance matrix of the measurements, in this
case specified by: $\{n,\,\s8,\,\om,\, w\}$. We have also introduced
the total number of bins $n_{\rm B}= n_{\rm B}^{\omega} + n_{\rm
  B}^{\Phi} + n_{\rm B}^{\gamma_T}$, as well as the total covariance
matrix of the measurements in bins $i$ and $j$:
\be C_{i j}=\langle (m_i-\bar m_i)\,(m_j-\bar
m_j)\rangle, \hspace{0.5cm} i,j=\overline{1, n_{\rm B}}
\label{eq:cov}
\ee
The Fisher matrix is defined as the ensemble-averaged Hessian of the
logarithm of the likelihood function,
\be \Fish_{\alpha \beta} = - \left< \frac{\partial^2 \ln {\cal
    L}}{\partial p_{\alpha}\,\partial p_{\beta}}\right>.
\label{eq:Fisher_gen}
\ee
Note that we shall also use \eqns{eq:cov}{eq:Fisher_gen} for each
probe individually, case in which the vectors $\m,\, \bar{\m}$ and
matrix $\C$ will implicitly contain only the measurements of the
respective probe, and $n_{\rm B}$ will be the number of bins for that
same probe. We shall make no further specification on this subject and
rely on the context for clarity.
\begin{figure*}
\centering {
  \includegraphics[scale=0.43]{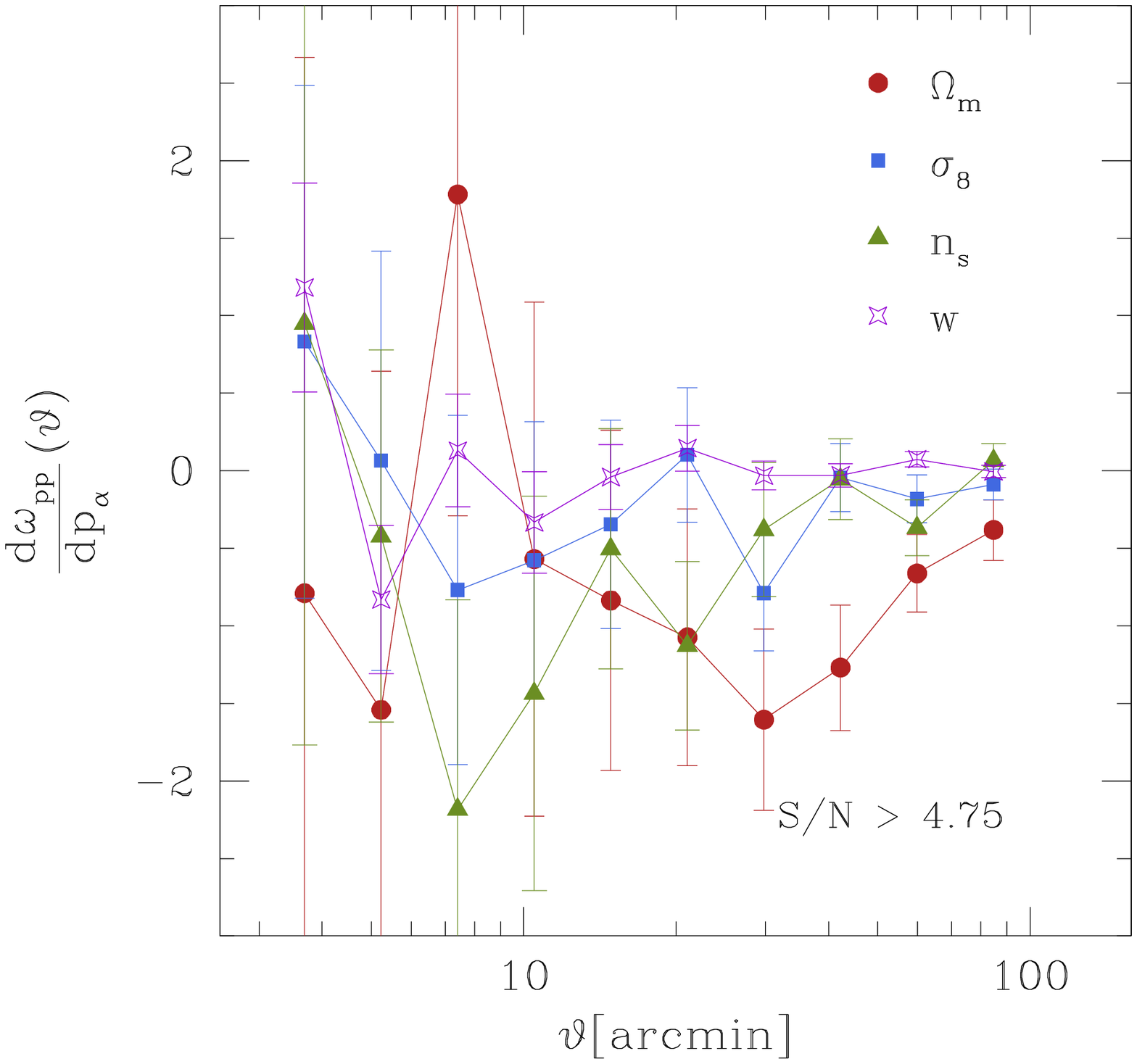}
  \includegraphics[scale=0.43]{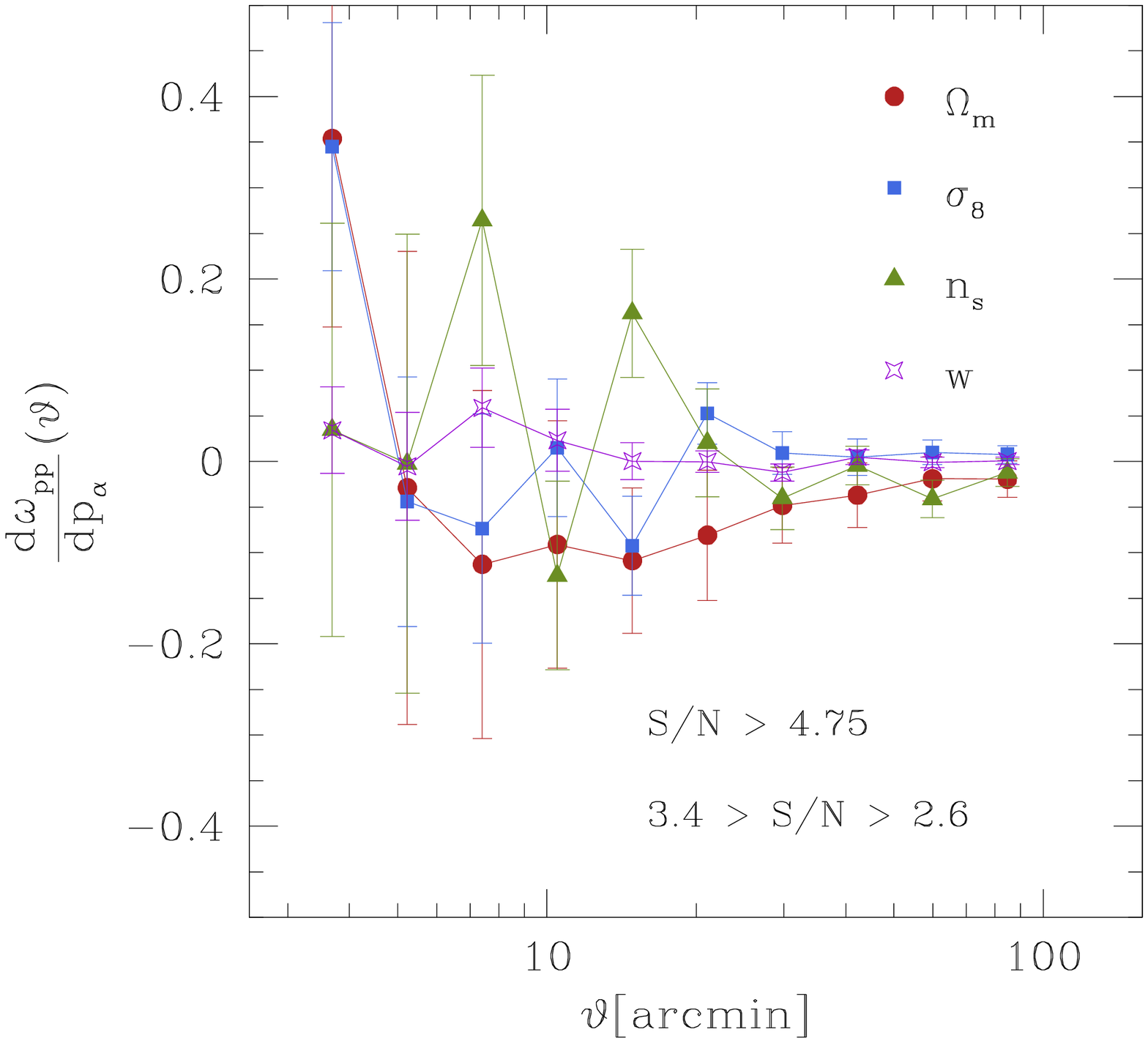}}
\caption {The derivatives of the peak correlation function measured
  from the noisy maps. The left panel shows the auto-correlation
  function $\corra$ for peaks with $\SN\geq 4.75$, while in the right
  one we present the cross-correlation $\corrc$ between two bins with
  low and high $\SN$.}
\label{fig:deriv_corr}
\end{figure*}
\begin{figure*}
\centering {
  \includegraphics[scale=0.43]{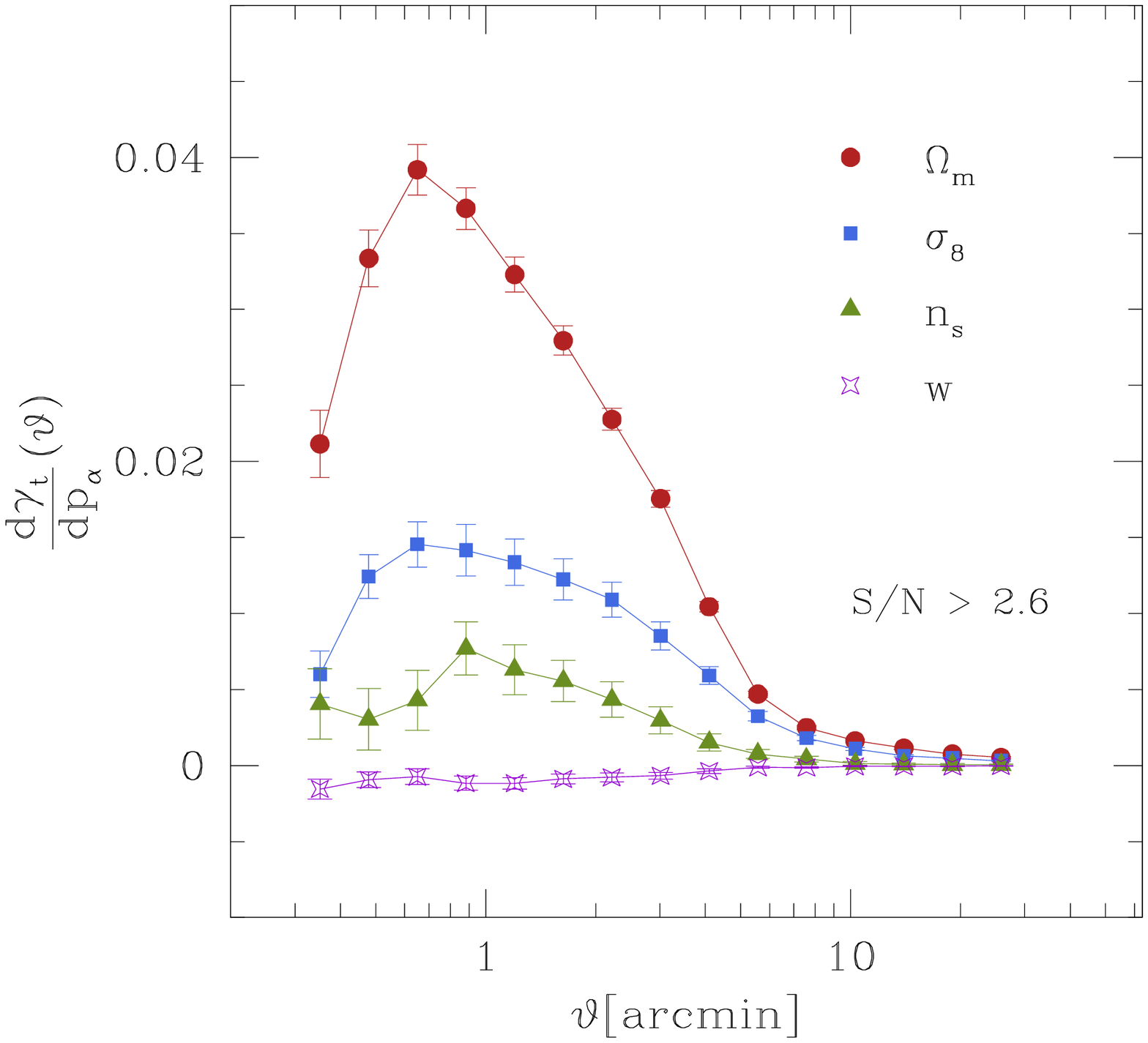}
  \includegraphics[scale=0.43]{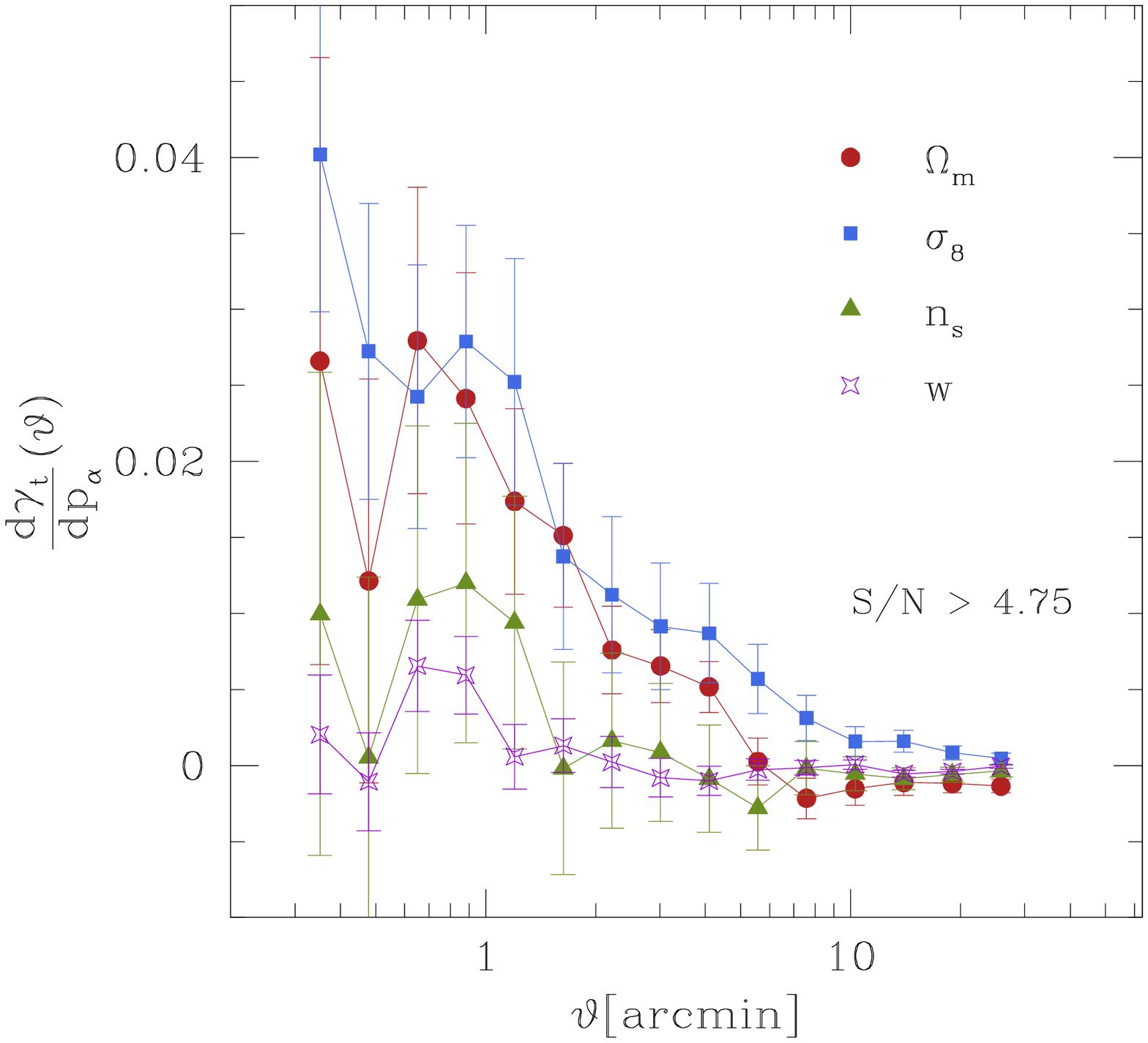}}
\caption {The derivatives of the tangential-shear profiles $\gamma_T$
  measured from the noisy maps. The left panel corresponds to the
  peaks detected with $\SN\geq2.6$, while in the right one, peaks have
  $\SN\geq4.75$.}
\label{fig:deriv_prof}
\end{figure*}

The general expression of \eqn{eq:Fisher_gen} can be rewritten as
\citep{Tegmarketal1997}
\be \Fish_{\alpha \beta} = \frac{1}{2} {\rm Tr}
\left[\C^{-1}\C_{,p_\alpha}\C^{-1}\C_{,p_\beta}\right] + \bar
     {\m}_{,p_\alpha}^{T} {\C}^{-1} \bar {\m}_{,p_\beta}
\label{eq:Fisher2}.
\ee
Here we have denoted by $x_{,p_\alpha}=\partial x/\partial
p_\alpha$. We shall ignore the first term in \Eqn{eq:Fisher2} for two
reasons: i) it is considered to contribute little to the Fisher
information -- for a discussion on this see \cite{Tegmarketal1997};
ii) an accurate determination of the derivatives of the covariance
with respect to the cosmological parameters would require more
realizations of the variational cosmologies than we currently
have. Therefore we shall follow the standard approach to Fisher-matrix
forecasting in the literature and ignore the trace-term in the above
equation.

From the Fisher matrix, one may obtain an estimate of the marginalized
errors and covariances of the parameters:
\be \sigma^2_{p_\alpha p_\beta} = [\Fish^{-1}]_{\alpha \beta},
\label{eq:marg_error}
\ee
as well as the unmarginalized errors:
\be \sigma_{p_\alpha} = [\Fish_{\alpha \alpha}]^{-1/2} .
\label{eq:unmarg_error}
\ee
%
We now turn our attention to the estimation of each element
contributing to the simplified expression of the Fisher matrix,
discussed above. For each cosmological model, the mean auto- and
cross-correlation functions are evaluated as an average of the
correlation functions for each field,
\be
\hat{\bar\omega}(\vartheta)=\frac{1}{n}\sum_{k=1}^{n} \hat{\omega}_k(\vartheta),
\label{eq:estmean}
\ee
where $\hat{\omega}_k$ is the correlation function measured from the
$k^{\rm th}$ field, and $n$ is the number of fields
considered. Similarly, we compute the mean of the peak function and
the tangential-shear profiles. An unbiased, maximum-likelihood
estimator for the covariance is
\be
\hat{C}_{i j}=\frac{1}{n-1}\sum_{k=1}^{n} (\hat{m}_i^k-\hat{\bar m}_i)
\,(\hat{m}_j^k-\hat{\bar m}_j).
\label{eq:cov2}
\ee
We use this estimator to compute the fiducial-model-covariance matrix
for each of the three probes, as well as for any combination of
them. The estimate of the inverse covariance is corrected in the
following way \citep[]{Anderson1958, Hartlapetal2007}:
\be
\widehat{\C^{-1}}=\frac{n-n_{\rm B}-2}{n-1}(\hat{\C})^{-1},\:  n_{\rm B} < n-2,
\label{eq:Cinv}
\ee
In order to obtain low-noise estimates of the derivatives with respect
to the cosmological parameters, we take advantage of the matched
initial conditions of the simulations, and use the double-sided
derivative estimator
\be
\widehat{\frac{\partial \bar m_i}{\partial p_\alpha}}=\frac{1}{n}\sum_{k=1}^{n} 
\frac{\hat{m}_i^k(p_\alpha +\Delta \alpha)-\hat{m}_i^k(p_\alpha-\Delta \alpha)}{2\Delta \alpha},
\label{eq:derivs}
\ee
where $\Delta \alpha$ represents the $\pm$ step in the cosmological
parameter $p_\alpha$, e.g. Table \ref{tab:zHORIZONcospar}. Note that
whilst this approach reduces the cosmic variance in the derivatives,
the estimated Fisher matrix is still noisy due to the inverse
covariance estimator.

Finally, we mention a technical point regarding the joined constraints
presented in the next section. Since we evaluate the combined
covariance for the three probes, we are concerned that its elements
might be very different, hence introducing numerical instabilities in
its inversion, and possibly leading to an inaccurate estimate of the
inverse. To prevent this, we apply the same strategy recently used in
\citep{Smithetal2012}: instead of using the covariance matrix, we use
the correlation matrix, defined by
\be
r_{i j}=C_{i j}/\sigma_i\,\sigma_j\ ,
\ee
where $\sigma_{i}=\sqrt{C_{i i}}$ is the rms variance. The inverse
correlation and covariance are related by:
\be
r^{-1}_{i j}=\sigma_{i} \sigma_{j} C^{-1}_{i j}
\ee
We can rewrite the Fisher matrix from \eqn{eq:Fisher2} as:
\be \Fish_{\alpha \beta} = \sum_{i, j=1}^{n_{\rm B}}\frac{\bar
  {m}_{i, p_\alpha}}{\sigma_i}\,{r}_{i j}^{-1}\, \frac{\bar
  {m}_{j, p_\beta}}{\sigma_j}
\label{eq:Fisher3},
\ee
It is this last equation that we shall be using for our forecast,
scaling the measured derivatives of the probes by the rms variance of
the fiducial functions, and employing the inverse correlation matrix,
instead of the inverse covariance.
\begin{figure}
\centering {
  \includegraphics[scale=0.44]{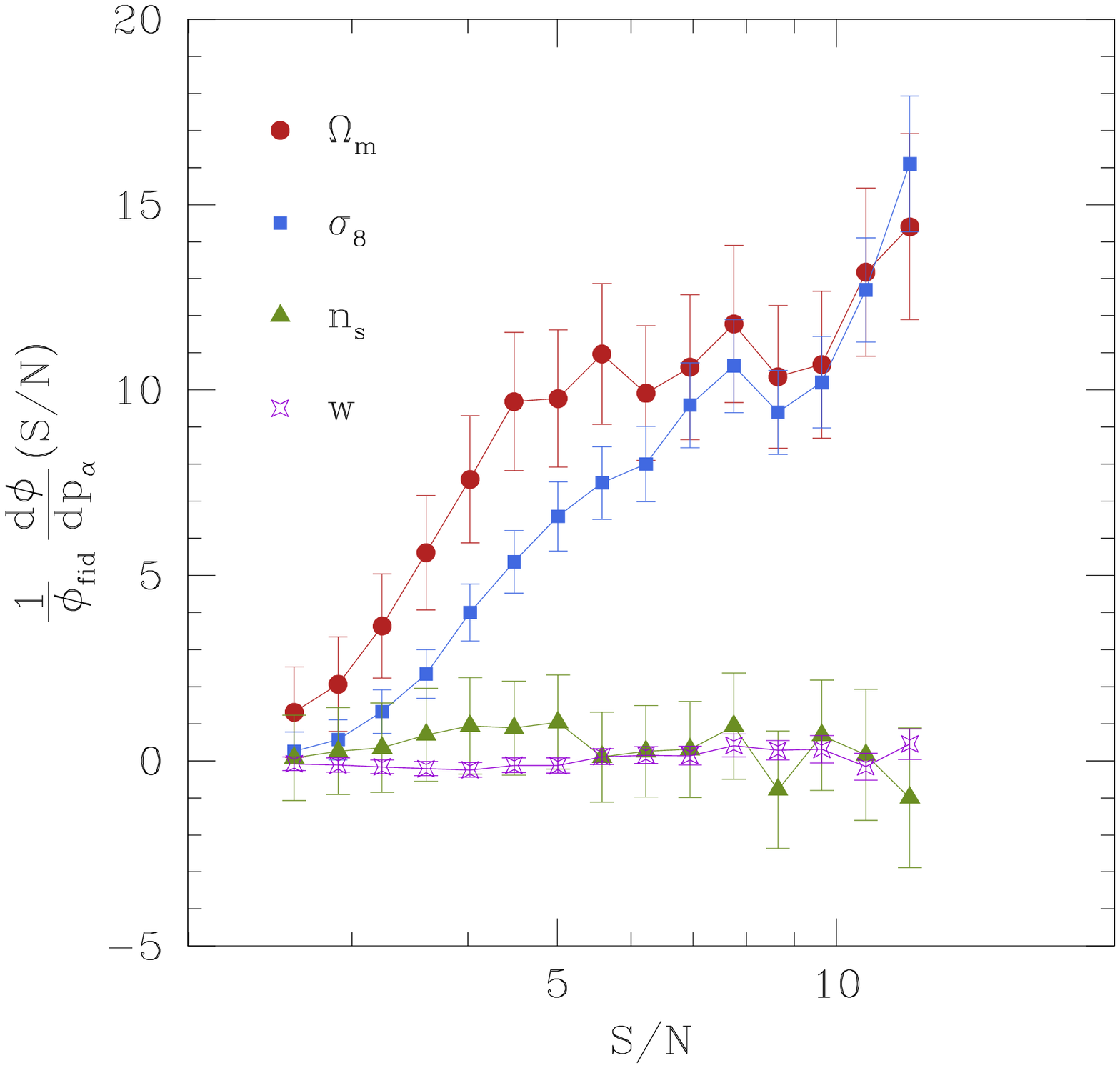}}
\caption {The derivatives of the peak function $\Phi$ measured from
  the noisy maps and scaled by the peak function corresponding to the
  fiducial cosmological model.}
\label{fig:deriv_counts}
\end{figure}
\fig{fig:deriv_corr} depicts the derivatives of the auto- and
cross-correlation function with respect to the cosmological parameters
that we investigate in this work -- left and right panels,
respectively. The auto-correlation derivatives are quite noisy --
primarily due to the relatively small number of peaks with $\SN\geq
4.75$, a characteristic also visible in \fig{fig:corrfunc}, the noise
being largest on small scales, $\vartheta < 20\,\arcmint$. The $\om$-
derivative has the largest amplitude, while $w$ has the smallest. All
derivatives tend to 0 on larger scales, signifying that there is no
information in the peak-peak correlation function for $\vartheta\geq
2\,{\rm deg}$. Compared to the auto-correlation function, the
derivatives of the cross-correlation are less noisy, with the
exception of $\ns$. They are also significantly smaller, just like the
cross-correlation signal is much smaller than the auto-correlation
one. In both panels, the $\om$-derivative is negative for
$\vartheta\geq 10\,\arcmint$, whilst for the other parameters, the
derivatives cross the 0-line a few times.

\fig{fig:deriv_prof} presents the shear-profile derivatives with
respect to the cosmological parameters: the left panel corresponds to
peaks with $\SN\geq 2.6$, while the right one is for $\SN\geq 4.75$,
hence the larger measurement noise present there. In both panels, the
derivatives decrease to 0 for increasing angular scales,
e.g. $\vartheta\sim20\,\arcmint$. In the left panel, all derivatives
are positive, with the only exception of $w$, which has a
barely-noticeable transition from slightly negative to slightly
positive. The $\om$-derivative is the largest, while the $w$-one is
the smallest. For the higher-$\SN$ peaks, the derivatives are less
smooth, but the general trends are preserved: $\om$ and $\s8$ still
have the highest derivatives, followed by $\ns$ and $w$. They all seem
to peak on the scales $0.8-1\,\arcmint$, which suggests that a
significant amount of information arises from these scales. We shall
mostly be using the larger peaks for our constraints.

\fig{fig:deriv_counts} depicts the derivatives of the peak function
with respect to the cosmological parameters, scaled by the peak
abundance corresponding to the fiducial cosmology, as a function of
$\SN$. Again $\om$ has the largest derivative, followed closely by
$\s8$ -- both of them positive. The derivatives with respect to $\ns$
and $w$ are much smaller and less smooth.

Since in this study we use numerical derivatives to estimate the
Fisher errors, as opposed to analytical ones, we need to ask what the
resulting uncertainty on the Fisher errors is. This is particularly
needed for our noisiest measurements, i.e. for the correlation
function. We address this question in the appendix \S\ref{A1}, and
refer the interested reader to that discussion.

One point of interest -- useful mainly to develop one's intuition --
is to explore which are the highest contributing scales to the Fisher
information. Naturally, we expect a combination of scales to be the
most effective at constraining the cosmological model; the most common
approach to determine such a combination is to diagonalize the
covariance matrix and determine its eigenvalues: the eigenvectors
corresponding to the largest eigenvalues will represent the most
constraining linear combinations of scales. To see if there is a
particular scale which helps to significantly reduce the errors, we
shall take two approaches. 
\begin{figure*}
\centering {
\includegraphics[scale=0.42]{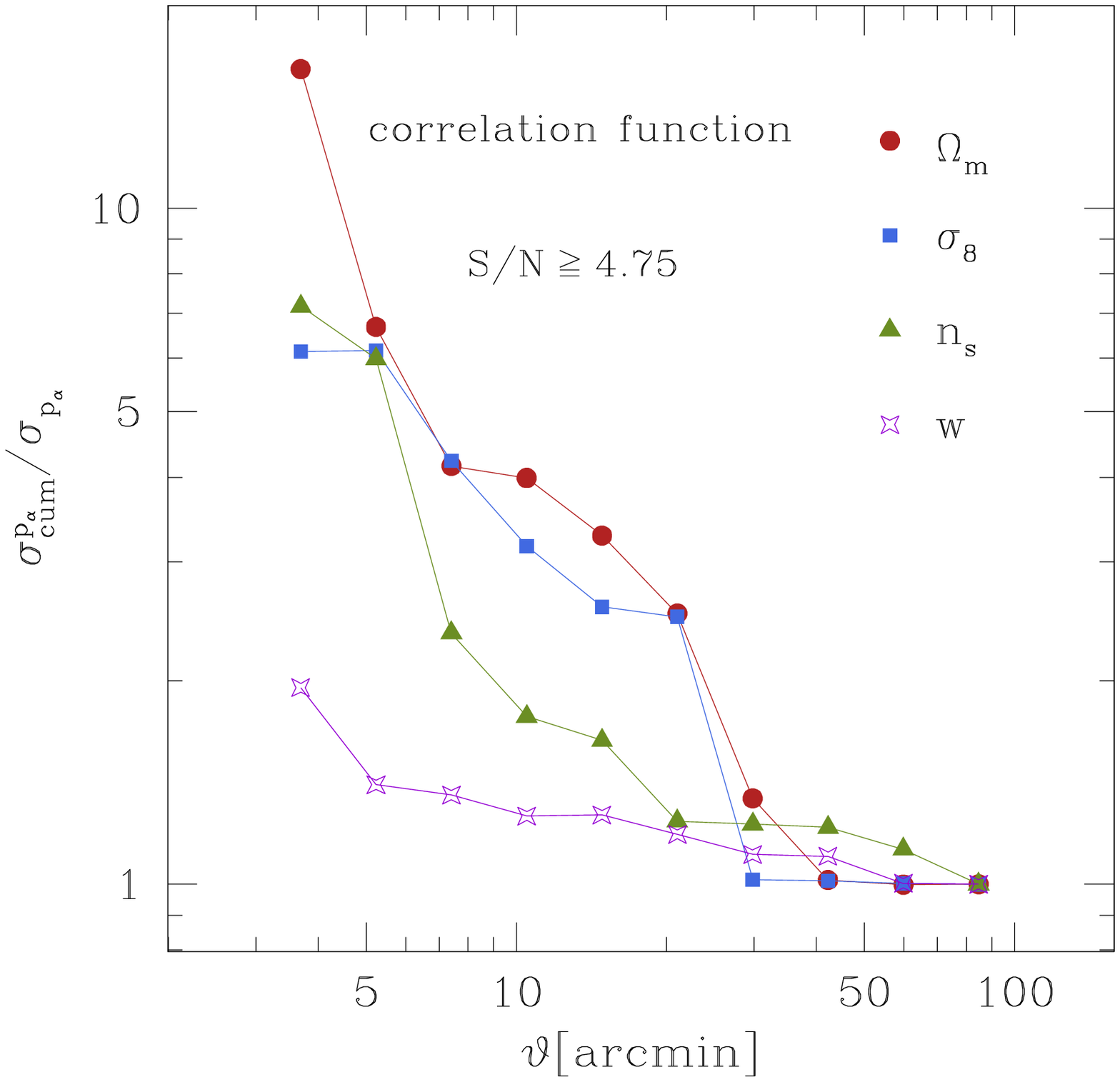}
\includegraphics[scale=0.42]{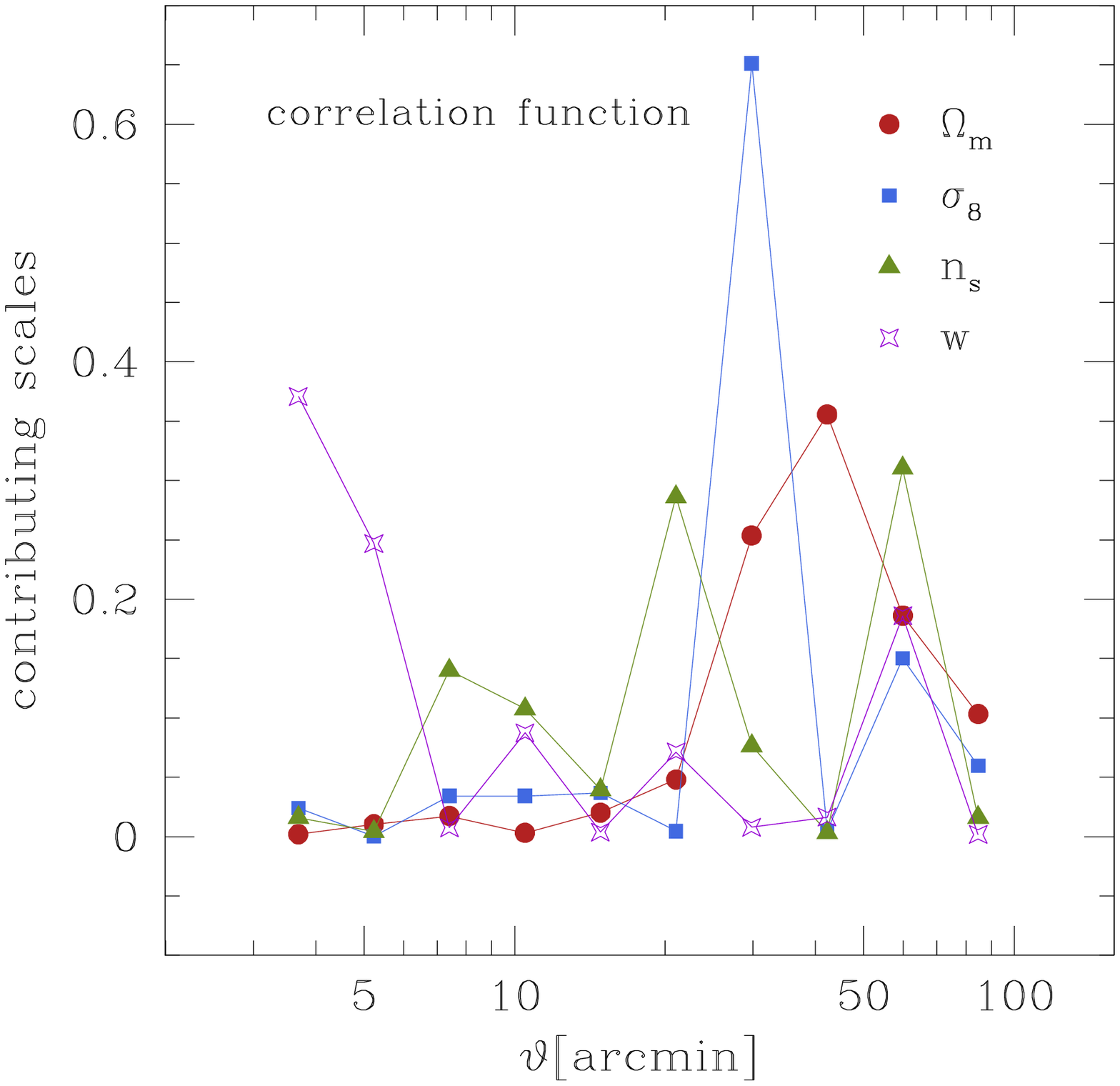}
\includegraphics[scale=0.42]{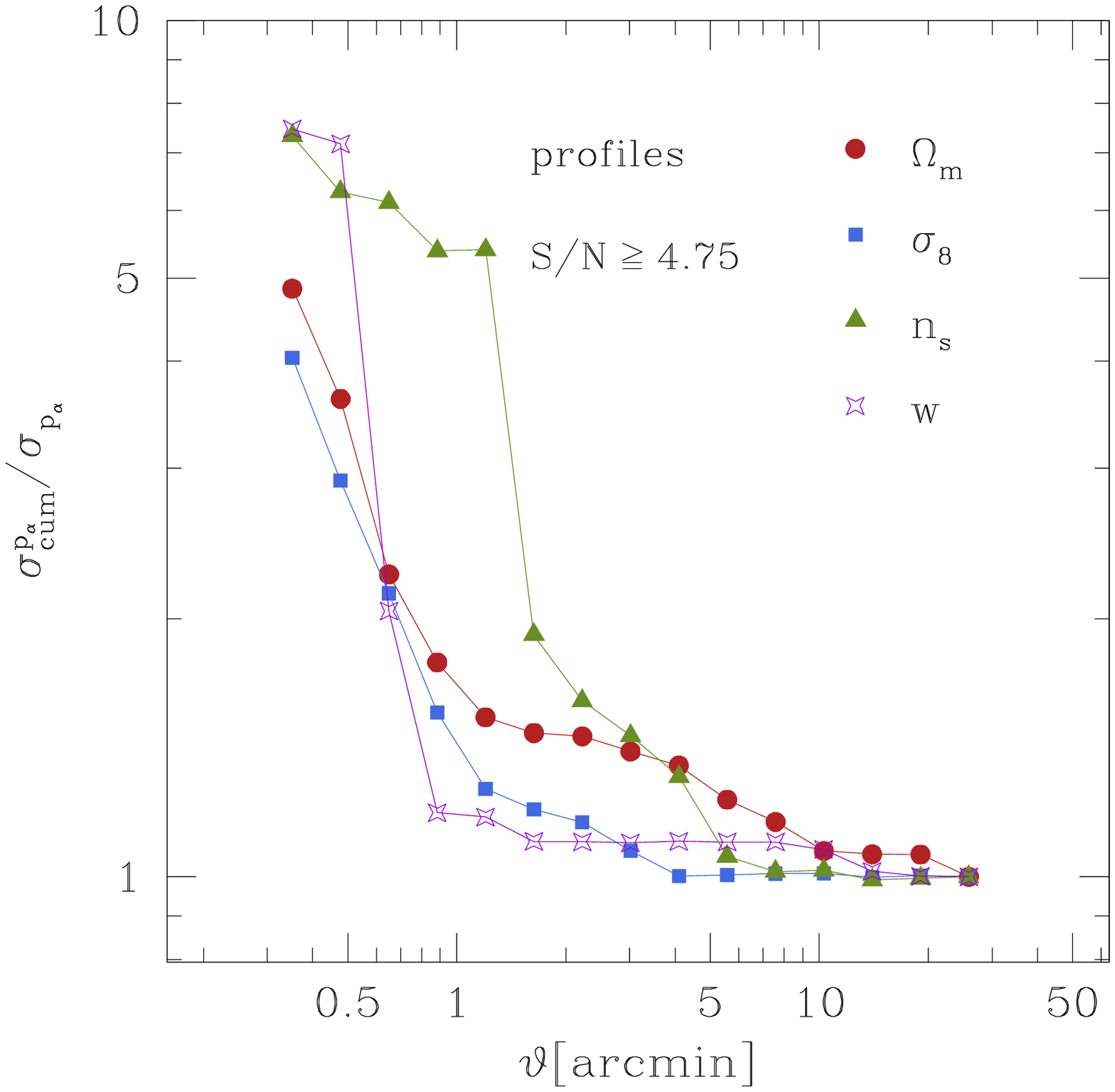}
\includegraphics[scale=0.42]{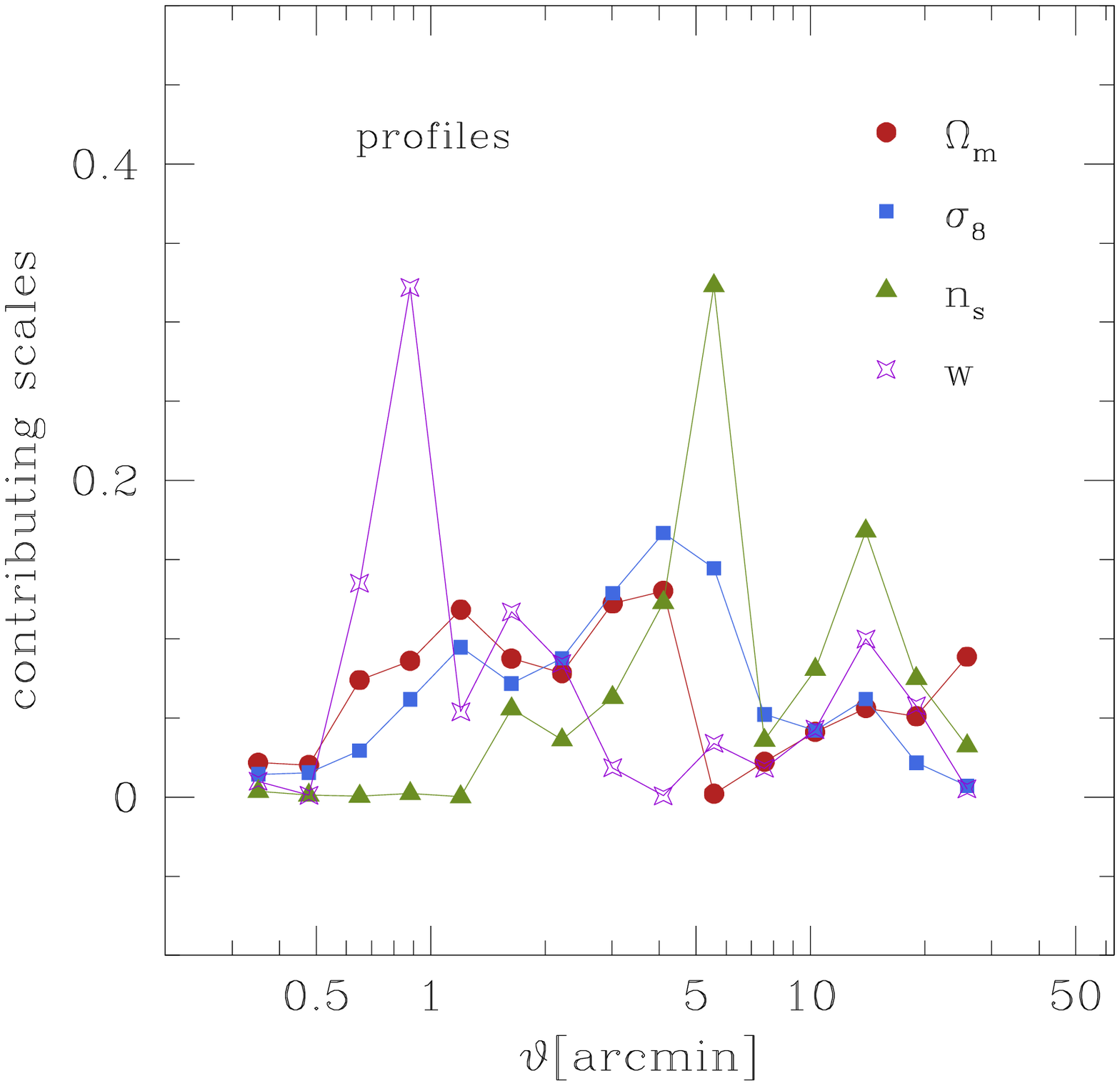}
\includegraphics[scale=0.42]{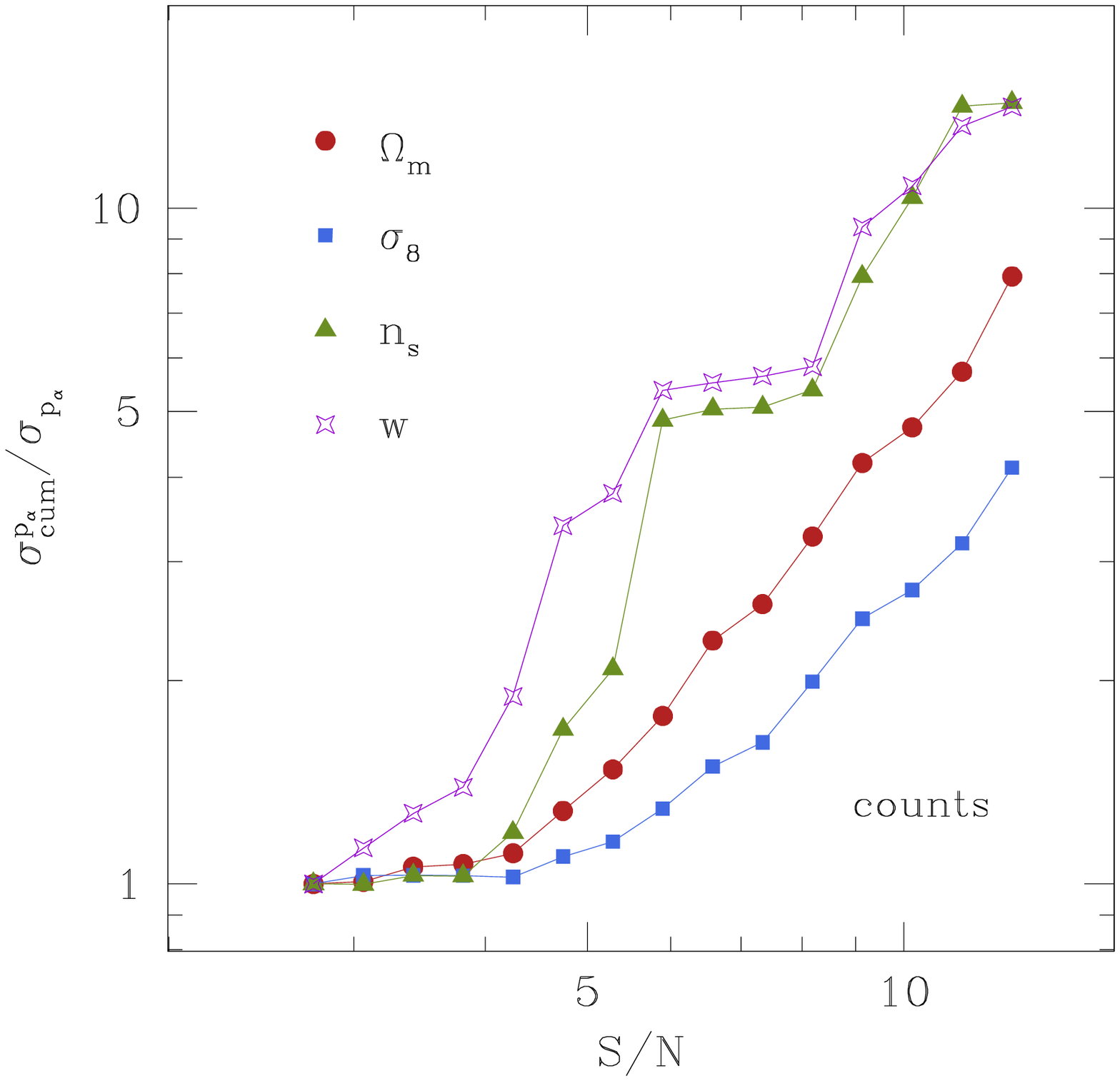}
\includegraphics[scale=0.42]{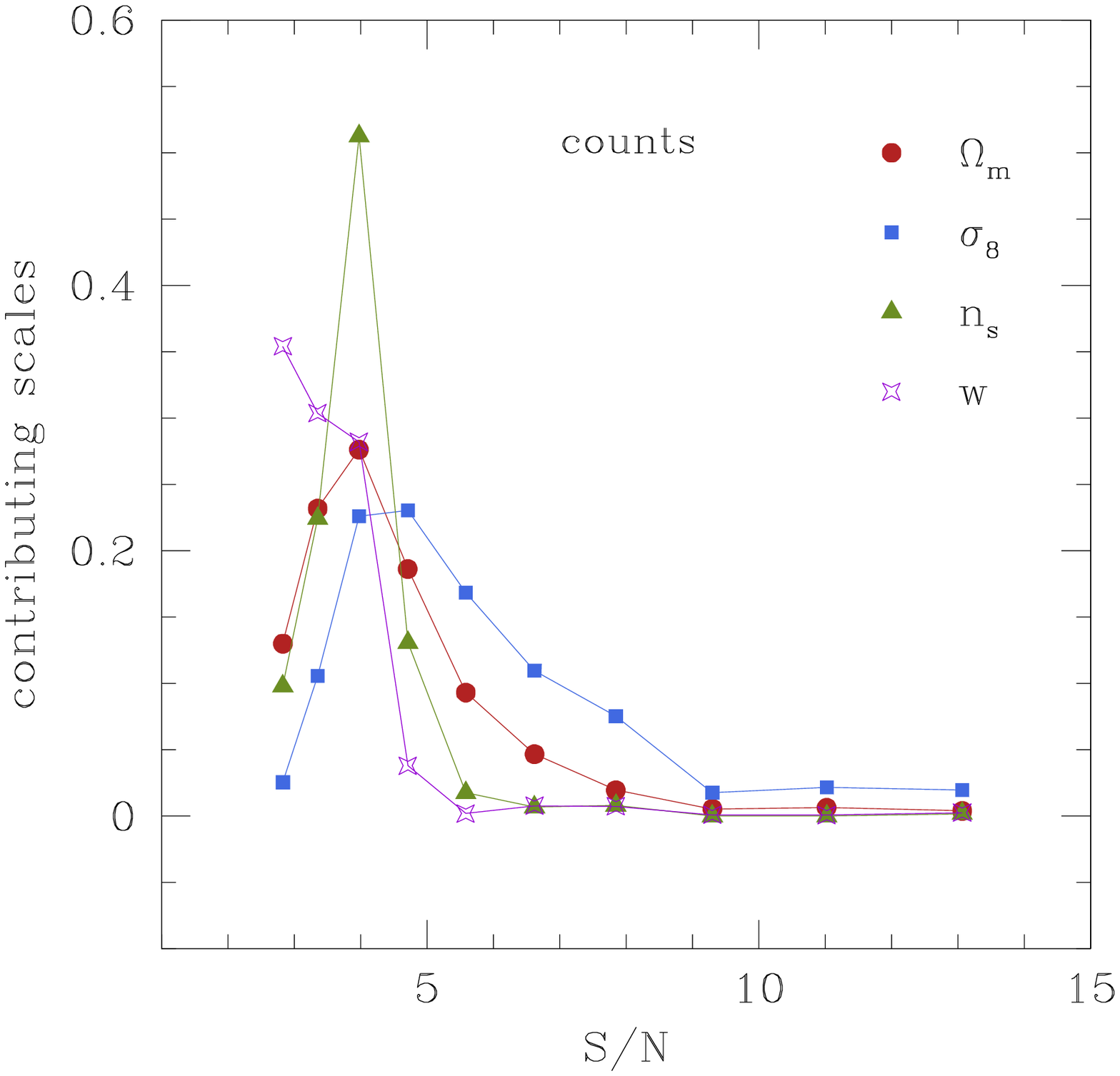}
}
\caption{{\it Left panels:} Cumulative unmarginalized Fisher errors
  for $\corra$, $\gamma_T$, and $\Phi$ -- top, middle, and bottom
  panels respectively. Bins are added progressively, from small to
  large scales for the angular bins, and from large to small
  $\SN$. The cumulative errors are scaled by the final error, i.e. the
  error obtained when all bins have been included. For the
  auto-correlation and the profiles, only peaks with $\SN \geq 4.75$
  are included. {\it Right panels:} The contribution of different
  scales to the Fisher information, estimated as the rms of
  \eqn{eq:FishDat2}.}
\label{fig:cum_scales}
\end{figure*}
First, we compute the cumulative Fisher matrix, i.e. we monitor how
the errors evolve when bins are included one-by-one in the calculation
of \Eqn{eq:Fisher2}. We choose as starting bins those containing
measurements expected to be easier to perform in real surveys. For
example, for the peak counts, we start with the highest-$\SN$ bins,
and then add progressively smaller-$\SN$ bins. For the profiles and
correlation functions, we start with small-scale measurements, and
progressively include large angular scales. The results of this
exercise are presented in the left panels of \fig{fig:cum_scales}:
from top to bottom, we show the cumulative unmarginalized Fisher
errors arising from $\corra,\,\gamma_{T},$ and $\Phi$, scaled by the
final errors -- when all bins are included -- as a function of
bins. For the auto-correlation function, the information seems to
saturate at $\sim 40\,\arcmint$: the inclusion of larger-scale
measurements does not further improve the constraints. For the
profiles, the saturation occurs at $\sim 10\,\arcmint$, while for the
peak counts, the inclusion of peaks with $\SN<4$ still improves the
constraint on $w$, though not on the other parameters. This is in line
with our findings from \cite{Marianetal2012}.
\begin{figure*}
\centering {
\includegraphics[scale=0.7, angle=-90]{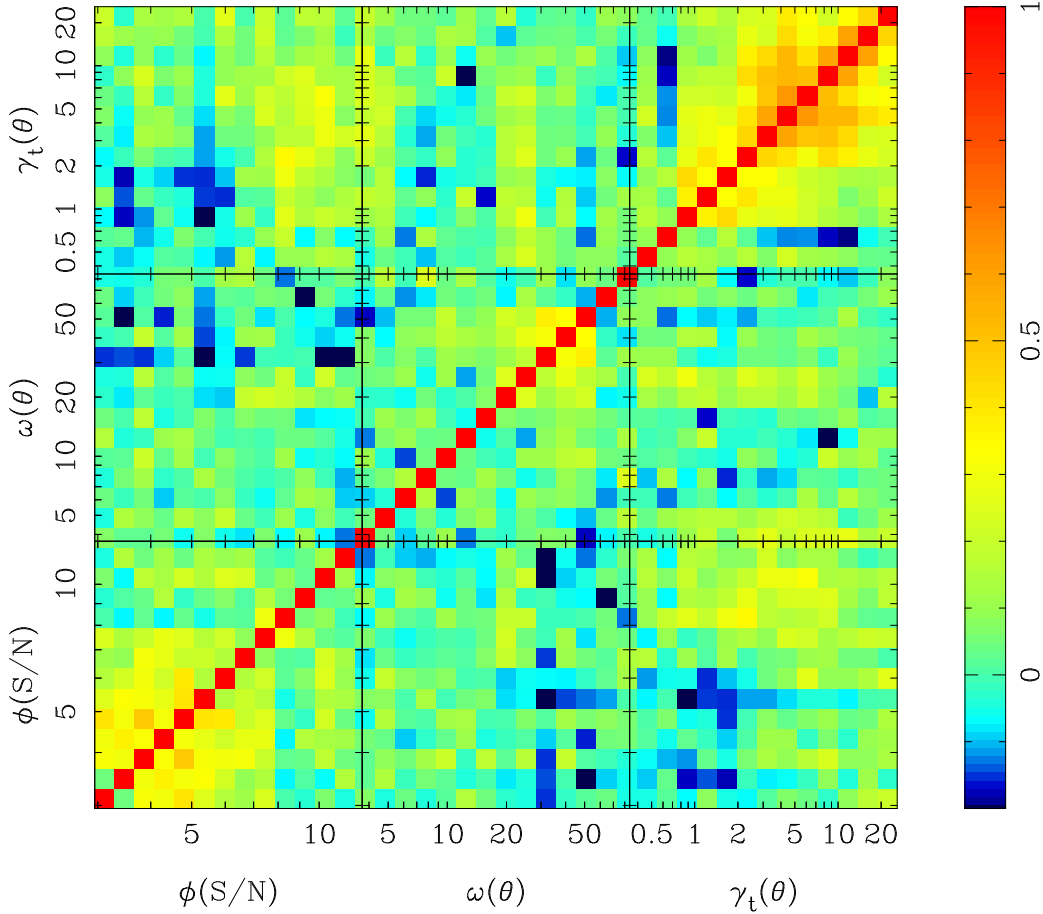}
\caption {Total cross-correlation matrix of the peak abundance $\Phi$,
  the peak-peak correlation function $\corra$, and the stacked peak
  profiles $\gamma_T$. $\corra$ and $\gamma_T$ are evaluated for peaks
  with $\SN\geq 4.75$, while $\Phi$ contains the peaks with
  $\SN\geq2.6$. From left-right and down-up, we plot the correlations
  of $\Phi,\,\corra,\,\gamma_T$, with the $\SN$ and angular scales
  (measured in $\arcmint$) increasing in the same directions. The
  correlation matrix is computed from 128 fields of
  $12\times12\,\deg2$ corresponding to the fiducial cosmology and is
  rescaled to match a sky coverage of $\sim18000\,\deg2$.}
\label{fig:cross_corr}}
\end{figure*}

The second approach involves two approximations: i) for each probe we
neglect the covariance of the parameters, i.e. we consider only
unmarginalized errors -- this was also done for the first
approach. ii) we disregard the off-diagonal elements of the covariance
matrix of each probe. The Fisher matrix can then be written as
\be 
\Fish_{\alpha \alpha} = \sum_{i=1}^{n_{\rm B}}\bar{m}_{i, p_\alpha}^2/C_{i i}.
\label{eq:FishDat}
\ee
We can quantify the contribution of each bin $j$ to the total Fisher
constraints through the ratio:
\be \frac{\Fish^{j}_{\alpha \alpha}}{\Fish_{\alpha
    \alpha}}=\frac{\bar{m}_{j, p_\alpha}^2/C_{j j}}{\sum_{i=1}^{n_{\rm
      B}}\bar{m}_{i, p_\alpha}^2/C_{i i}}.
\label{eq:FishDat2} 
\ee
By construction, this ratio works best as a contributing-scales
indicator for those probes which have a covariance matrix with as few
off-diagonal elements as possible. In our case, this is the
 auto-correlation function, as can be seen from
\fig{fig:cross_corr}. The rms of the above ratio is shown in the right
panels of \fig{fig:cum_scales}, ordered in the same way as the left
ones. For the correlation function, the most contributing scales to
the $\s8$- and $\om$-constraints are $\sim 30$ and $\sim 40\,\arcmint$
respectively. This is in agreement with the left panel of the figure:
the errors on those two parameters are lowered significantly when the
respective bins are included in the Fisher estimates. The constraints
from stacked profiles receive contributions from all scales below
$10\,\arcmint$, while for the counts the smaller bins ($\SN\leq 5$)
tighten the errors -- though notice that the agreement between the
left and right panel is not so good as for the correlation function,
due to the high correlations between the low-$\SN$ bins. These
correlations are included in the left panel, but not in the right one.

\section{Cosmological constraints}
\label{S5}
\begin{figure*}
\centering {
\includegraphics[scale=0.7]{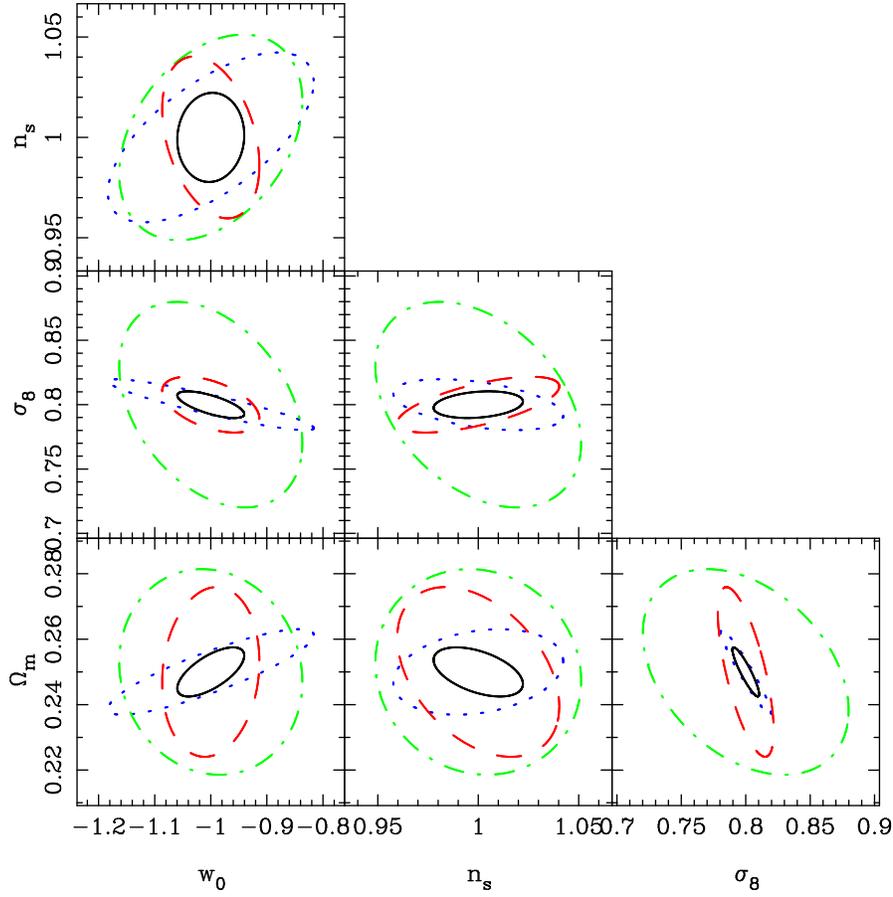}
\caption {2-$\sigma$ Fisher ellipses for the peak probes. The
  auto-correlation function is depicted by the green dot-dashed
  ellipse; the red dashed ellipse represents the profiles, while the
  dotted blue line corresponds to the peak abundance. The combination
  of the three is given by the black ellipse.}
\label{fig:ellipses}}
\end{figure*}
\begin{figure*}
\centering {
\includegraphics[scale=0.7]{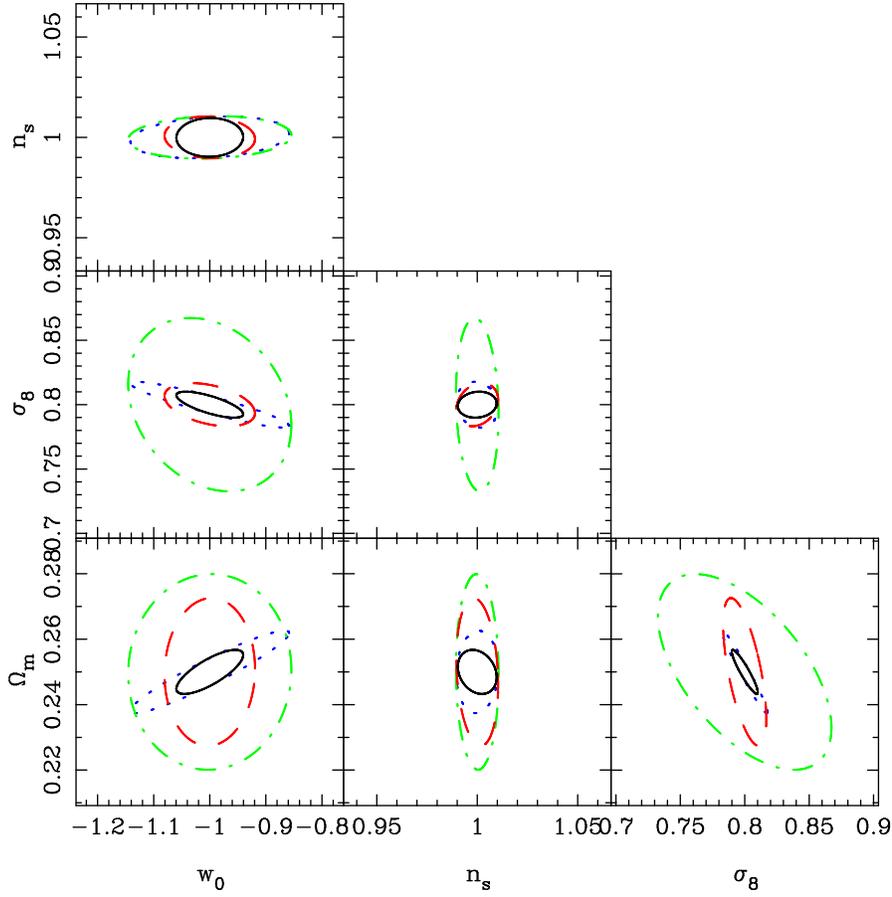}
\caption {2-$\sigma$ Fisher ellipses for the peak probes, with a
  $\planck$-like prior. The colors and symbols are the same as in the
  previous figure. The prior has been added to each peak probe
  separately.}
\label{fig:ellipses_P}}
\end{figure*}
Using all the ingredients described in section~\S\ref{S4}, we now
proceed to constraining the cosmological model using the three peak
probes individually, as well as in combination. To this avail, we
shall employ only noisy maps, and consider the auto-correlation
$\corra$ of peaks with $\SN\geq 4.75$, the cross-correlation $\corrc$
of peaks with $\SN\geq 4.75$ and $3.4\geq\SN\geq2.6$, the stacked
profiles $\gamma_T$ of peaks with $\SN\geq4.75$, and the abundance of
peaks $\Phi$ with $\SN\geq 2.6$. The Fisher matrix errors are
estimated using the covariance on the mean of the fiducial model for
the three probes of interest to us here. Thus, our estimated
covariance from \eqn{eq:cov2}, which corresponds to an area of
$144\,\deg2$, is rescaled to correspond to an area 128 times larger,
i.e. $\sim 18000\,\deg2$. Together with the survey specifications
given in section \S\ref{S2}, this makes our study representative for
two future surveys, $\lsst$ and $\euclid$.

\fig{fig:cross_corr} presents the cross-correlation matrix ${\bf r}$
of these probes. By far, $\gamma_T$ has the strongest correlation
coefficient of the three: $\sim 0.7$ on scales $2-20\, \arcmint$. For
the peak function, the low-$\SN$ bins are the most correlated $\sim
0.5$ for $\SN\leq 5$. This was already established in our earlier work
\citep{Marianetal2012}, and it can be explained through the
better-known behaviour of halos: small-mass halos are sample-variance
dominated, while the large and rare halos follow the Poisson
distribution \citep{HuKravtsov2003, SmithMarian2011}. Note however
how the smallest-$\SN$ bins in \fig{fig:cross_corr} seem to be
completely uncorrelated: this is most likely due to the overwhelming
number of shape-noise peaks, which are random, unclustered, and
therefore uncorrelated. $\corra$ displays the smallest correlation
coefficient of the three, $\sim 0.3-0.4$ on the scales
$20-60\,\arcmint$, with weaker correlations on smaller scales. We
further note the weak cross-correlation of $\corra$ and $\Phi$, as well
as $\corra$ and $\gamma_T$. There is a visible cross-correlation of
$\Phi$ and $\gamma_T$, of $\sim 0.3$ for peaks with $\SN>7$. This is
most likely due to the stacked profiles being dominated by the most
massive peaks, which also dominate the high-$\SN$ end of the peak
function.
\begin{table*}
\caption{Fisher-matrix constraints using $\corra,\,\corrc$,
  $\gamma_T$, and $\Phi$. All errors are presented as percentages of
  the fiducial values of the parameters,
  ${\om=0.25,\,\s8=0.8,\,w=-1,\,\ns=1}$.
\label{tab:Fisher_constraints}}
\vspace{0.2cm}
\centering{
\begin{tabular} {c}
{\bf Unmarginalized errors [\%] } \\
\begin{tabular}{c|ccccccccccc}
\hline 
 &  $\corra$ & $\corrc$ & $\gamma_T$ & $\Phi$ & $\corra + \Phi$ & $\corra + \gamma_T$ & $\corrc + \Phi$ & $\corrc + \gamma_T$ & $\Phi + \gamma_T$ & $\corra + \Phi + \gamma_T$ & $\corrc + \Phi + \gamma_T$ \\
{\tt $\om$}  & 3.7 \ &  6.7 & 3.3 & 0.3 & 0.3 & 2.4 & 0.3 & 3 & 0.3 & 0.3 & 0.3  \\
{\tt $\s8$}  & 3.2 \ &  3.4 & 1.8  & 0.2  & 0.2  & 1.4  & 0.2  & 1.6 & 0.2 & 0.2 & 0.2 \\
{\tt $w$}    & 5.6 \ &  10.2  & 3.5  & 1.6  & 1.5   & 3  & 1.5   & 3.2 & 1.3 & 1.3 & 1.4 \\
{\tt $\ns$}  & 1.9 \ &  1.7 & 1.6  & 0.7  & 0.6  & 1.1  & 0.7  & 1.1 & 0.6 & 0.6 & 0.6 \\
\hline
\end{tabular} \\
{\bf Marginalized errors [\%] } \\
\begin{tabular}{c|ccccccccccc}
\hline 
 &  $\corra$ & $\corrc$ & $\gamma_T$ & $\Phi$ & $\corra + \Phi$ & $\corra + \gamma_T$ & $\corrc + \Phi$ & $\corrc + \gamma_T$ & $\Phi + \gamma_T$ & $\corra + \Phi + \gamma_T$ & $\corrc + \Phi + \gamma_T$ \\
{\tt $\om$}  & 4.4 \ & 6.8 & 4  & 1.9 & 1.4 & 3 & 1.5  & 3.7   & 1.2 & 1.1  & 1.1 \\
{\tt $\s8$}  & 4.1 \ & 4 & 2  & 1 & 0.7 & 1.8 & 0.7 & 1.8 & 0.6 & 0.5 & 0.5 \\
{\tt $w$}    & 6.4 \ & 11.5 & 4  & 6.7  & 4.5 & 3.2 & 4.5 & 3.4 & 2.9  & 2.7 & 2.7  \\
{\tt $\ns$}  & 2.1  \ & 2.1 & 2  & 1.6 & 1.2 & 1.2 & 1.1 & 1.3 & 1 & 0.8 & 0.9  \\
\hline
\end{tabular}
\end{tabular}}
\label{fisher_er}
\end{table*}
Table~\ref{fisher_er} presents the unmarginalized and marginalized
1-$\sigma$ errors resulting from the three peak probes. Each probe
taken by itself, the abundance of peaks has the greatest constraining
power, followed by the profiles, and then by the correlation function.
Regarding the latter, we note that $\corra$ and $\corrc$ yield very
similar constraints, the auto-correlation being more effective for $w$
and $\om$ -- a reduction by factors of $\sim 2$ and $\sim 1.5$
respectively in these errors, compared to the
cross-correlation. However, when combined with the other two probes,
there is little difference between $\corra$ and $\corrc$.  The
greatest benefit to adding the correlation function or the profiles to
the abundance of peaks concerns the time-independent equation-of-state
for dark energy: after marginalizing over the other parameters, the
errors on $w$ resulting from $\Phi$ and $\corra$ taken individually
are similar, while the profiles seem to yield a constraint tighter by
a factor of $\sim 1.7$. When all three probes are combined, the
constraints on $\om,\,\s8,\,\ns$ improve by a factor of $\sim 1.5-2$
compared to using $\Phi$ alone, while for $w$ the improvement is $\sim
2.5$. Lastly, combining $\Phi$ and $\gamma_T$ is almost as efficient
as using all three probes: the contribution of the correlation
functions to reducing the errors is negligible, if both the abundance
and the profiles are used. 

This is further detailed by \fig{fig:ellipses} which shows the
forecasted likelihood contours, at 95\% confidence level. The largest
ellipses -- green, dot-dashed -- represent the $\corra$ constraints;
for many of the parameter combinations, the red, dashed ellipses
depicting $\gamma_T$ are almost perpendicularly oriented relative to
the $\corra$ ellipses, and also quite tilted with respect to the
$\Phi$ ellipses, shown by the blue dotted lines. The greatest gain in
combining all three probes is for the case of $w$, where the resulting
ellipse is smaller by a factor of $\sim 3$ compared to the $\Phi$
ellipse.

We compare our Fisher constraints to those of \cite{Hilbertetal2012},
derived for an identical survey scenario from the same simulated WL
maps. That analysis combines the shear correlation functions, the
third moment of the aperture mass, and the peak abundance, and
accounts also for local primordial non-Gaussianity. The percentage
unmarginalized combined 1-$\sigma$ errors found in
\cite{Hilbertetal2012} for $\{\om,\,\s8,\,w,\,\ns\}$ were
$\{0.2,\,0.15,\,0.9,\,0.4\}$. The marginalized errors were
$\{0.7,\,0.4,\,2.6,\,0.7\}$. Their constraints are tighter,
particularly for $\om$, but the combination of peak probes performs
nonetheless very well. 

\fig{fig:ellipses_P} presents the 95\%-likelihood contours of the
three probes, after we have added cosmic microwave background (CMB)
information resulting from an experiment similar to $\planck$. For the
latter, we assume that the CMB temperature and polarization spectra
can constrain 9 parameters: the dark energy equation-of-state
parameters $w_0$ and $w_a$; the density parameter for dark energy
$\Omega_{\rm DE}$; the CDM and baryon density parameters scaled by the
square of the dimensionless Hubble parameter $\omega_{\rm
  CDM}=\Omega_{\rm CDM}\,h^2$ and $\omega_{\rm b}=\Omega_{\rm b}\,h^2$
($h=H_0/[100\, {\rm km\,s^{-1}\,Mpc^{-1}}]$); the primordial spectral
index of scalar perturbations $\ns$; the primordial amplitude of
scalar perturbations $A_s$; the running of the spectral index
$\alpha$; and the optical depth to the last scattering surface
$\tau$. We compute the CMB Fisher matrix as described by
\cite{Eisensteinetal1999}:
\be 
\Fish_{\alpha \beta}=\sum_l \sum_{X,Y} \frac{\partial C_{l,X}}{\partial p_{\alpha}}
{\rm Cov}^{-1}\left[C_{l,X},C_{l,Y}\right]
\frac{\partial C_{l,Y}}{\partial p_{\beta}}\ ,
\ee
where $\{X,Y\}\in \{{\rm TT},\,{\rm EE},\,{\rm TE},\,{\rm BB}\}$,
where $C_{l,\rm TT}$ is the temperature power spectrum, $C_{l,\rm EE}$
is the E-mode polarization power spectrum, $C_{l,\rm TE}$ is the
temperature-E-mode polarization cross-power spectrum, and $C_{l,\rm
  BB}$ is the B-mode polarization power spectrum. The assumed sky
coverage is $f_{\rm sky}=0.8$ In order to make the CMB Fisher matrix
compatible with our parameters, we rotate it to a new set
$
{\bf q}^{T}=\{w_0, w_a, \Omega_{m}, h, f_b, \tau, \ns, \s8, \alpha \} \ ,
$
where for us $w_0=w$. We marginalize over the 5 parameters absent from
our analysis, and then add the resulting $4\times4$ matrix to the
Fisher matrix for each of the probes, and their combination. This way
of implementing the $\planck$ prior follows the same steps taken in
our previous studies \citep{Marianetal2012, Hilbertetal2012}.
\fig{fig:ellipses_P} shows that combining the peak probes with the CMB
dramatically alters all the ellipses involving $\ns$: they are
shrunken and almost completely aligned. This is hardly surprising, as
the CMB constrains $\ns$ much better than any WL probe, so the
ellipses are then dominated by the CMB information. For the other
parameters, the changes are small, and there is still a significant
gain to be obtained by combining the peak probes, compared to using
only one of them, for instance the abundance.
\section{Summary and conclusion}
\label{S6}
In this paper we have investigated cosmological constraints from WL
peaks, beyond using just counts. We have measured for the first time
the peak-peak correlation function, and estimated its constraining
power on the following cosmological parameters:
$\{\om,\,\s8,\,w,\,\ns\}$. We have also measured the tangential-shear
profiles of the peaks, and explored their cosmological utility. Whilst
both the correlation function and stacked profiles of
galaxies/clusters are standard tools for cosmology, our study applies
them to shear-selected objects, detected only through their
gravitational lensing effects. We employed mostly numerical methods,
performing measurements of the peak abundance, peak profiles, and
peak-peak correlation functions in simulated WL maps. To estimate the
cosmological constraints, we used the Fisher-matrix formalism,
computing the derivatives with respect to cosmological parameters and
the combined covariance of the three peak probes from the simulated
maps. The latter have been generated with ray-tracing through a large
suite of $N$-body simulations, varying the cosmological model around
some fiducial values. Given the distribution of source galaxies and
the level of shape noise considered, our simulated maps are relevant
for future large WL missions, such as $\euclid$ and $\lsst$.

To measure the correlation function of peaks, we used the
\cite{LandySzalay1993} estimator, known for two essential properties:
i) minimum variance; ii) lack of bias due to the unknown mean density
of the measured objects. We have shown that the correlation function
has its maximum on scales $\vartheta\leq 10\,\arcmint$, with the
position and height of the latter depending on the $\SN$ of the
considered peaks. For angular scales $\vartheta\geq 8\,\arcmint$, the
correlation function of `large' peaks ($\SN\geq 4.75$) has a greater
amplitude than that of `medium' and `small' peaks ($4.75>\SN\geq 3.4$
and $\SN<3.4$), mirroring the clustering behaviour of dark-matter
halos. We have shown that the low-$\SN$ end of the peak function is
completely dominated by shape-noise peaks -- only $\sim5\%$ of peaks
with $\SN\sim 3$ are genuine LSS peaks. This is the reason why the
correlation function of the small peaks quickly declines to $0$:
shape-noise peaks are random, and therefore uncorrelated. On the
contrary, the correlation function of the peaks measured from
noise-free maps is quite high, even with all peaks, small and large,
binned together: this is just a statement that peaks due genuinely to
projected LSS are quite clustered. The cross-correlation of small- and
large-$\SN$ peaks measured from noisy maps is also smaller than its
noise-free counterpart, but it can still be used for cosmological
constraints.

We have measured the tangential-shear profiles of the peaks using
their WL centre, as detected by our hierarchical method. We found that
the larger the $\SN$ of the peaks, the higher their average profile. A
comparison of the results from noise-free and noisy maps showed that:
i) the ratio of the average profile of high-$\SN$ peaks to the average
profile of all peaks combined together is $\sim 2$ for noise-free
maps, and $\sim 5$ for noisy maps; ii) the profile average of all
noisy peaks binned together is a factor of $\sim 5$ smaller than that
of the noiseless profiles; iii) the profiles of genuine LSS peaks are
shallower on scales $\vartheta<2\,\arcmint$ when the peaks are
detected and measured from noisy maps, rather than noise-free maps --
this owes to a coordinate shift of the peak centres in the presence of
shape noise. The first two findings are mostly due to the overwhelming
number of shape-noise peaks which dominate the low-$\SN$ end of the
peak function: measurements of the tangential-shear profiles around
random points in the map bring down the average of the stacked
profiles of genuine LSS structures.

The Fisher-matrix analysis that we performed revealed the forecasted
errors on the cosmological parameters from the auto- and
cross-correlation functions of the peaks, as well as from the stacked
profiles to be larger by at least a factor of $2$ than those obtained
using the peak function. Nevertheless, combining the peak function
with any of the other three probes reduces the errors significantly,
particularly for $w$. The ratio of the marginalized 1-$\sigma$
constraints from $\Phi$ alone to the constraints from $\corra + \Phi$
is $\{1.36,\,1.43,\,1.5,\,1.3\}$, corresponding to
$\{\om,\,\s8,\,w,\,\ns\}$; for $\gamma_T + \Phi$ the ratio is
$\{1.6,\,1.7,\,2.3,\,1.6\}$; and for $\corra + \gamma_T + \Phi$, the
result is $\{1.7,\,2,\,2.5,\,2\}$.  Therefore, one of the main
conclusions of this study is that future WL surveys should use peak
statistics beyond the 1-point function. Note that our $\corra +
\gamma_T + \Phi$ constraints are quite competitive with those derived
by \cite{Hilbertetal2012} from combining the shear correlation
functions, the third moment of the aperture mass, and the peak
abundance. For the parameters that we have investigated, the stacked
tangential-shear profiles seem slightly more constraining than the
correlation function. However, if other parameters were to be included
in the analysis, particularly those quantifying primordial
non-Gaussianities, the peak-peak correlation function might be more
powerful. This however, is the subject of an ongoing work, to be
presented in the near future.

We have briefly examined the $\SN$ and angular scales contributing
most to the Fisher information. The cosmology constraints from the
peak-peak correlation function benefit significantly from measurements
at scales $\vartheta\sim 20-40\,\arcmint$, and saturate for
$\vartheta>50\,\arcmint$. For the tangential-shear profiles,
measurements at $\vartheta\sim1-10\,\arcmint$ are important, and the
constraints saturate for $\vartheta>10\,\arcmint$. The peak abundance
benefits from the inclusion of small-$\SN$ bins.

In this study, we have not exhaustively investigated the properties of
the clustering of the peaks. Measuring the correlation function was
only the first step of such a study. It would also be interesting to
examine the biasing of the peaks with respect to the shear/convergence
field, and also to compare the clustering of the peaks with that of
the halos in the simulations. Our goal for the near-future is to
determine the full constraining power of the shear-selected WL peaks
on an expanded set of cosmological parameters, i.e. to extend the
recent study of \cite{Hilbertetal2012} to include the 2-point
statistics of the peaks, as well as tomographic techniques.
\subsection*{Acknowledgements}
We are grateful to the Institute for Theoretical Physics of the
University of Z\"urich for its hospitality. We thank V. Springel for
making public {\tt Gadget-2} and for providing his B-FoF halo
finder. LM and PS are supported by the Deutsche Forschungsgemeinschaft
(DFG) through the grant MA 4967/1-2, through the Priority Programme
1177 `Galaxy Evolution' (SCHN 342/6 and WH 6/3), and through the
Transregio TR33 `The Dark Universe'. SH acknowledges support by the
National Science Foundation (NSF) grant number AST-0807458-002.

\bibliographystyle{mn2e} 

\appendix
\section{Estimating the error on the Fisher constraints}
\label{A1}
In Fisher forecasts as well as real-data analysis, the derivatives of
observables with respect to cosmological parameters are usually
estimated analytically. However, due to the lack of analytical
predictions for peak observables, in this study we employ numerical
derivatives. As discussed in \S\ref{S4}, the latter are estimated as
double-sided derivatives (\eqn{eq:derivs}), which reduces the impact
of cosmic variance and the dependence on the step $\Delta p_{\alpha}$
in the cosmological parameters around the fiducial value. One natural
question is: what is the impact of the uncertainties in the measured
derivatives on the resulting Fisher constraints? We shall address this
issue in an approximate manner, just for the unmarginalized errors. 

The unmarginalized Fisher constraints are computed from
\eqns{eq:unmarg_error}{eq:Fisher3}. Their variance can be written as
\be {\rm Var}[\sigma_{p_{\alpha}}]= \frac{1}{4}\Fish_{\alpha
  \alpha}^{-3} \,{\rm Var}[\Fish_{\alpha \alpha}].  \ee
Since we are interested in the impact of the uncertainties in the
derivative estimates on the Fisher matrix, we ignore the uncertainties
in the covariance matrix of the probes. In any case, we would not be
able to reliably estimate the latter from just 128 realizations, so
this approximation is quite necessary. Note that the impact of
uncertainties in the covariance matrix on Fisher constraints was
explored in the recent study of \cite{Tayloretal2012}. The variance of
the Fisher-matrix elements can then be written as
\ba 
{\rm Var}[\Fish_{\alpha \beta}] = \sum_{i, j=1}^{n_{\rm B}}\left(\frac{r_{i j}^{-1}}
{\sigma_i \sigma_j}\right)^2 \, \{\bar{m}_{j, p_\beta}^2 {\rm Var}
[\bar{m}_{i, p_\alpha}] \nn \hspace{1cm}\\ + \bar{m}_{i, p_\alpha}^2 {\rm Var}
[\bar{m}_{j, p_\beta}] +2\,\bar{m}_{i, p_\alpha}\bar{m}_{j, p_\beta}{\rm Cov}
[\bar{m}_{i, p_\alpha}\bar{m}_{j, p_\beta}]\}.
\label{eq:fishvar}
\ea
The rms of the variance of the derivatives represents the error bars
shown in \figss{fig:deriv_corr}{fig:deriv_prof}{fig:deriv_counts}. The
covariance in the above equation must be estimated separately from the
data. Since we have only 64 realizations corresponding to each of the
variational cosmologies, this estimate is bound to be very
noisy. Whilst $\bar{\m}$ can be any of the three peak observables, we
actually perform the above calculation for the correlation function
alone. The derivatives of the latter have the largest error bars and
they also float around the 0 line, which could yield spurious Fisher
information. Here we mention that increasing the number of bins to 12
and 15 did not significantly change our parameter constraints for
$\corra$. The uncertainties in the unmarginalized 1-$\sigma$ Fisher
constraints for $\corra$, expressed as percentages of the values given
in Table~\ref{fisher_er} are for $\{\om,\,\s8,\,w,\,\ns\}$:
$\{26\%,\,31\%,\,17\%,\,26\%\}$ respectively -- if we ignore the noisy
covariance term in \eqn{eq:fishvar}; and
$\{38\%,\,43\%,\,24\%,\,36\%\}$ -- if we do take into account the
covariance term.

We leave to a future work a more complete treatment of the
Fisher-matrix constraints from numerical measurements. We would like
to analytically model the peak probes so as not to be vulnerable to
numerical effects when computing their derivatives with respect to
cosmological parameters. 

\end{document}